\documentclass[journal]{IEEEtran}

\usepackage[pdftex]{graphicx}

\usepackage{amssymb,amsmath}

\usepackage{bbold}
\usepackage{hyperref}
\hypersetup{hidelinks}

\usepackage{tabularx}
\usepackage{comment}
\usepackage[dvipsnames]{xcolor}
\usepackage{booktabs}
\usepackage{array}
\usepackage{dcolumn}
\usepackage{makecell}
\usepackage{multirow}
\usepackage[labelfont=bf]{caption}
\usepackage{amsthm}
\usepackage{mathtools}
\theoremstyle{definition}
\newtheorem{definition}{Definition}[section]
\usepackage{cleveref}
\Crefname{figure}{Fig.}{Figs.}
\Crefname{equation}{Eq.}{Eqs.}
\Crefname{section}{Sec.}{Secs.}

\usepackage{pifont}
\newcommand{\cmark}{\ding{51}}
\newcommand{\xmark}{\ding{55}}

\usepackage[noadjust]{cite}
\usepackage[numbers,sort&compress]{natbib}

\ifCLASSINFOpdf

\else

\fi

\ifCLASSOPTIONcompsoc
\usepackage[caption=false,font=normalsize,labelfont=sf,textfont=sf]{subfig}
\else
\usepackage[caption=false,font=footnotesize]{subfig}
\fi

\begin{document}

\title{Cell-based VSC Analysis Methodology: From Graph Laplacian to Converter Degrees of Freedom}

\author{Daniele~Falchi,~\IEEEmembership{}
        Eduardo~Prieto-Araujo,~\IEEEmembership{Senior Member,~IEEE,}
        and~Oriol~Gomis-Bellmunt,~\IEEEmembership{Fellow,~IEEE}
\vspace{-4mm}}

\maketitle

\begin{abstract}
Power-electronics-based converters are being considerably employed through the power system to interconnect multiple heterogeneous electrical layers. Furthermore, the intrinsic versatility to play with the converter network topology is widely exploited to accommodate a certain number of terminals and ports according with the specific application. On this regard, several converter arrangements can be encountered in power applications. 
Moreover, to properly establish both the operation and the control, the so-called degrees of freedom (DOFs) need to be assessed per each converter topology. On this matter, similarly to the well-known Clarke transformation, which clearly reveals the DOFs for the star-based topology system, further similar transformations can be achieved to depict the independent set of variables characterizing a certain converter structure. 
Referring to the cell-based class of Voltage Source Converter (VSC) topologies, including Modular Multilevel Converter (MMC); this article proposes a general methodology to determine the change of variable matrix transformation for several converter arrangements which are related to complete bi-partite and multi-partite graphs. The methodology lies in the graph Laplacian spectral analysis, which remarks the structural normal modes at the converter points of connections. Furthermore, for a complete characterization, the instantaneous power patterns formulations, based on the DOFs, are also introduced.
\end{abstract}

\vspace{-3mm}
\IEEEpeerreviewmaketitle
\section{Introduction}
\IEEEPARstart{T}{he} recent electric power system developments are channelled to contain heterogeneous AC-DC technologies. The augmented interconnection among different systems voltage levels, frequencies and pole arrangements, is aided by Power Electronics (PE) based energy conversion stages \cite{wang2013harmonizing,chi2020circuits}.

\begin{figure}[!h]
	\centering
	\subfloat[]
	{
		\includegraphics[]{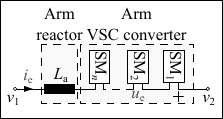}
		\label{fig:real_edge}
	}
	\subfloat[]
	{
		\includegraphics[]{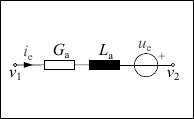}
		\label{fig:modelled_edge}
	}\\
	\vspace*{-0.8em}
	\subfloat[]
	{
		\includegraphics[]{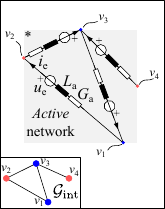}
		\label{fig:ga}
	}
	\hspace*{-0.6em}
	\subfloat[]
	{
		\includegraphics[]{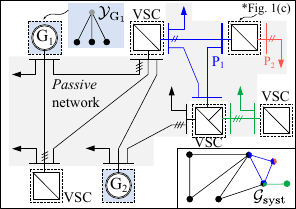}
		\label{fig:gb}
	}
	\caption{The arm converter schematic in (a) and its lumped-element equivalent model in (b). In (c), the converter \textit{active} network with its graph representation $\mathcal{G}_\mathrm{int}$. In (d), an arbitrary power system with its \textit{passive} network graph  $\mathcal{G}_\mathrm{syst}$.}
	\label{fig:general_system}
\end{figure}

Referring to high voltage applications, the state-of-the-art solution lies in the cell-based converter concept. Its essentials were introduced in \cite{baker1975electric}; as depicted in \Cref{fig:real_edge}, the so-called arm converter is constituted by a series interconnection of several Voltage Source Converter (VSC) sub-modules (SM), such that the internal SM capacitor can be either inserted or by-passed to establish the desired arm voltage across terminals $\text{\textit{v}}_1 \text{ and }\text{\textit{v}}_2$. In its reduced lumped-element model version, \Cref{fig:modelled_edge}, the arm converter is synthesized by an ideal controllable voltage source $\text{\textit{u}}_\text{e}$, and the series connection with the conductance component $G_\mathrm{a}$ of the arm reactor $L_\mathrm{a}$ form the equivalent Thevenin's representation. The concept is absolutely versatile; on one side, different voltage levels, without semiconductor ratings alteration, can be accommodated between terminals by playing with the number \textit{n} of SMs. On the other side, by properly assembling several arm converters to each other, further points of connections to multiple external systems are allowed. The specific arrangement of the arms establish the so-called converter topology which can be also considered as an \textit{active} network. In \Cref{fig:ga}, an example of an arbitrary cell-based arrangement composed by four arms is illustrated. From the circuit theory nomenclature, maintaining the port-terminals definitions presented in \cite{willems2010terminals,willems2013power}, the four available converter terminals $\text{\textit{v}}_1$, $\text{\textit{v}}_2$, $\text{\textit{v}}_3$ and $\text{\textit{v}}_4$, are grouped to form two ports, $\textit{P}_{1}=\left\lbrace \text{\textit{v}}_1,\text{\textit{v}}_3\right\rbrace $ and $\textit{P}_2=\left\lbrace \text{\textit{v}}_2,\text{\textit{v}}_4\right\rbrace $,  and used to transfer power among the two systems, as depicted in the arbitrary power system representation of \Cref{fig:gb}.

With these premises, in terms of converter circuit topologies, the original cell-based converter concept in \cite{baker1975electric} was further extended to obtain the so-called Modular Multilevel Converter (MMC) \cite{lesnicar2003innovative} and, nowadays, this VSC based technology is widely considered either for High Voltage Direct Current (HVDC) \cite{francos2012inelfe} or STATCOM \cite{peng2004universal} applications. Moreover, the cell-based converter concept is considerably spreading to other possible uses, e.g., DC/DC interconnections \cite{paez2018overview}, Static Frequency Converter (SFC) \cite{zhang2022design}, multi-port converters \cite{rouhani2019multiport,ma2021mmc} and Power Electronic Transformer (PET) \cite{ma2019multiport}.
Therefore, due to the increased variety of converter configurations, a systematic methodology analysis might be beneficial to comprehend the fundamental behaviour and properties, such as the degrees of freedom (DOFs) and inherent terminals-port affinities. In this regard, since the study of network topologies is commonly assessed by theory of graphs and linear algebraic tools \cite{maeda1957topological,seshu1959linear}; this article aims to introduce such a perspective to characterize the different converter topologies properties.
\vspace{-0mm}

In fact, since the G. Kirchhoff's work on resistive networks \cite{kirchhoff2003solution}, several works on multi-port transformer-less \textit{passive} networks employed a graph-based approach to establish a fundamental rule behind the coupling-decoupling among the ports of a certain topology and, eventually, build-up resistive network topologies with the desired properties \cite{brown2003synthesis,lupo2003synthesis,cederbaum1962realization,thulasiraman1969modified}. The fundamental insights are revealed by the specific admittance matrices which basically describes the interconnectivity properties between the nodes of the network. Recently, the application of graph theory was extended to power systems (\textit{passive} networks) with the principle objective to assess the network modelling and its controllability between terminals, generators and loads, \cite{ishizaki2018graph,doria2016modelling}. Along this line, as introduced in \cite{thulasiraman2018network}, the structural characteristics of a network and its impact on the dynamical behaviour of the interconnected systems might be revealed by the topological properties. In this regard, alternative works propose to tackle the topic from the corresponding graph Laplacian (Kirchhoff matrix) spectral, or eigenmode, analysis including graph reduction approaches and the effective resistance concept \cite{dorfler2012kron,song2019extension,zhu2021average,amani2023discovering}.

However, compared with the above mentioned \textit{passive} networks, the cell-based converter topologies deserve a further analysis. From KCL and KVL in \cite{maeda1957topological}, since the internal converter topology is defined as an \textit{active} network, the current through the terminals-edges cannot be established by the external sources only. Under the hypothesis of linearity, the overall behaviour would result from a linear superposition effect of both the internal and external sources. Furthermore, from the spectral point of view, preserving the linearity, as illustrated in \cite{endo1976mode,endo1976mode2} for LC multi-mode oscillators networks, a circuit of many DOFs have many independent modes,
the sum of these modes would be the solution of the original system by the principle of superposition. Additionally, contrarily to the \textit{passive} networks, in an \textit{active} network, the edges current may exist despite the terminals currents are zero.

\begin{figure}[!h]
	\centering
	\subfloat[]
	{
		\includegraphics[]{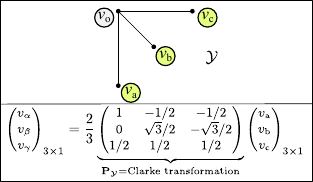}
		\label{fig:general_transf_1}
	}
	\hspace*{-0.8em}
	\subfloat[]
	{
		\includegraphics[]{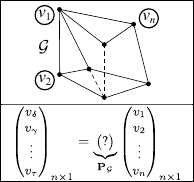}
		\label{fig:general_transf_2}
	}
	\caption{The Clarke transformation in (a). An unknown transition matrix for an arbitrary $n$-node graph  topology $\mathcal{G}$ in (b). }
	\label{fig:general_transf}
\end{figure}

Then, assuming that PE non-linearities do not play fundamental roles in establishing the spectral properties of a certain \textit{active} network topology, the article intends to propose a systematic approach to find the DOFs and characterize the affinities between terminals and ports of a given converter. In this direction, one of the initial works on the electrical DOFs of an \textit{active} network was introduced on the first half of the 20th century, when the electric power system was predominantly established by three-phase AC alternators, transformers and loads. E. Clarke, through the so-called $\alpha,\beta,0$ transformation \cite{clarke1943circuit}, advanced a linear transition matrix, \Cref{fig:general_transf_1}, between the terminals variables of the typical star-based topology ($\mathcal{Y}_\mathrm{G_1}$ in \Cref{fig:ga}) and its corresponding normal modes. This transformation, together with R. H. Park transformation \cite{park2009two}, are nowadays employed to analyse and control several applications based on the $\mathcal{Y}$ topology. 
Later on, the extension of a star-based topology to \textit{n}-phases \cite{parsa2005advantages,levi2008multiphase} leaded to revisit the original Clarke transformation matrix into a generalized form presented in \cite{rockhill2015generalized}. E. Clarke works  might be considered as pioneer in the field of spectral analysis; in fact, the transformation is actually revealing the typical star-based structure normal modes which can be valid either for the electrical or the mechanical system.

From a practical perspective, the converter topology spectral analysis, by decomposing the overall behaviour into multiple independent normal modes, it leads to a deeper understanding on the converter variables. This is relevant for both the design and operation. In fact, control-wise speaking, it fits very well with the so-called modal control strategy \cite{simon1968theory}. This control technique relies on controlling the overall system behaviour by regulating each independent normal mode; this is very suitable for multi-variable systems such as the cell-based converter topologies where multiple edges-nodes currents should be regulated at the desired value.

Moreover, from the nodal normal modes analysis, it is possible to assess the terminals-ports affinities per each mode and then characterize inherent port, or interconnected systems, coupling-decoupling. In this regard, it is found that the most common cell-based converter topologies lie in the complete bi-partite (and multi-partite) graphs families, \cite{bondy1976graph,chartrand2019chromatic}; these categories leads to precise inherent topological properties on the disjoint terminals subsets.

Through this manuscript, a further analysis on the instantaneous power variables description across the converter topology (terminals and edges) is proposed. Typically, referring to the \textit{passive} nature of circuits, the ports power analysis is mostly considered \cite{willems2013power,barbi2021theorem,barbi2024instantaneous}. On the other side, exploiting the graph Laplacian spectral analysis, it is worth to notice the work \cite{retiere2019spectral}, which illustrates a power flow along transmission lines. In this regard, the article aims to extend the Akagi's instantaneous power analysis, proposed in \cite{akagi2017instantaneous}, to different topologies in order to assess the energy behaviour from the edges-nodes voltage-current DOFs. Below, the main contributions are summarized:
\begin{itemize}
	\item A systematic methodology to identify both the nodal normal modes and the internal independent loops, for several cell-based converter topologies $\mathcal{G}$, \Cref{fig:general_transf_2}, is proposed.
	\item A systematic methodology to extract the instantaneous power DOFs of different topologies, from the decoupled voltage-current normal modes knowledge, is proposed. 
	\item From the topology nodal normal modes, inherent affinities among the terminals are disclosed. This in found to be fundamental to assess the eventual inherent topology decoupled ports, which is relevant when multiple systems are interconnected via converter terminals.
	\item Referring to the complete \textit{k}-partite based graphs, additional insights on possible alternative multi-port converter topologies are also disclosed. 
\end{itemize}
\vspace{-6mm}

The remainder of this paper is organized as follows. \Cref{sec:method_section} introduces the proposed methodology to assess overall converter topologies DOFs; with this purpose, the fundamental elements from graph, circuit and linear algebra fields used in the analysis, are presented. \Cref{sec:application} exposes the procedure on several complete \textit{k}-partite based converter topologies. Then, in \Cref{sec:insights}, further elements related to the presented methodology are disclosed; such as the opportunity to look for alternative multi-port converter topologies and inherent port decoupling structure categories of topologies. The conclusion in \Cref{sec:conclusion}. 
\section{Graph Laplacian Based Methodology}\label{sec:method_section}
\subsection{Complete $k$-partite Graphs}
\begin{figure}[!h]
	\centering
	\subfloat[]
	{
	\includegraphics[]{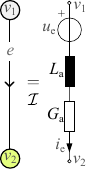}
	\label{fig:I_graph}
	}
	\hspace*{2mm}
	\subfloat[]
	{
	\includegraphics[]{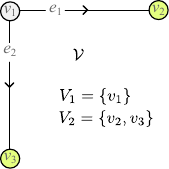}
	\label{fig:V_graph}
	}
	\hspace*{2mm}
	\subfloat[]
	{
	\includegraphics[]{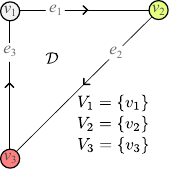}
	\label{fig:Delta_graph}
	}\\
	\vspace*{-2mm}	
	\subfloat[]
	{
	\includegraphics[]{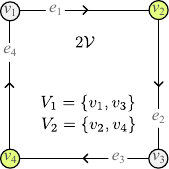}
	\label{fig:2V_graph}
	}
	\hspace*{-0.9em}
	\subfloat[]
	{
	\includegraphics[]{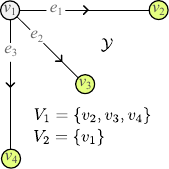}
	\label{fig:Y_graph}
	}
	\hspace*{-0.9em}
	\subfloat[]
	{
	\includegraphics[]{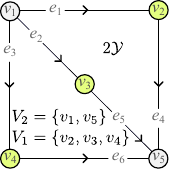}
	\label{fig:2Y_graph}
	}
	\caption{A complete bi-partite graph $K_{1,1}$ ($\mathcal{I}$) with the equivalent arm converter model representation in (a). A complete bi-partite graph $K_{1,2}$ ($\mathcal{V}$ topology) in (c) and a complete three-partite graph $K_{1,1,1}$ ($\mathcal{D}$) in (d). A complete bi-partite graph $K_{2,2}$ (2$\mathcal{V}$) in (d); a complete bi-partite graph $K_{3,1}$ ($\mathcal{Y}$) in (e) and, a complete bi-partite graph $K_{3,2}$ (2$\mathcal{Y}$) in (f). }
	\label{fig:common_topologies}
\end{figure}

As mentioned in the introduction, within the several families used to categorized graphs, there are the so-called complete bi-partite (and multi-partite) graphs, \cite{bondy1976graph} and \cite{chartrand2019chromatic}, which are found to be largely used for building up cell-based converters. Let $\mathcal{G}\left(V_\mathcal{G},E_\mathcal{G} \right) $ be a graph with $n$ vertices $V_\mathcal{G} = (v_1 , v_2,..., v_n)$ and $m$ edges $E_\mathcal{G} = (e_1 , e_2,..., e_m)$; referring to \Cref{fig:common_topologies}, their graph-wise definitions are introduced to highlight some inherent features exploited, in practice, on their corresponding electrical converter sub-networks. 
\begin{definition}[Complete bi-partite graph]
	A complete bipartite graph is a graph whose vertices $V_\mathcal{G}$ can be partitioned into two subsets $V_1$ and $V_2$ such that no edge has both endpoints in the same subset, and every possible edge that could connect vertices in different subsets is part of the graph (\Cref{fig:I_graph}, \Cref{fig:V_graph}, \Cref{fig:2V_graph}, \Cref{fig:Y_graph} and \Cref{fig:2Y_graph}).
\end{definition}
\begin{definition}[Complete multi-partite graph]
	A complete $k$-partite graph is a $k$-partite graph ($i.e.$, a set of graph vertices $V_\mathcal{G}$ decomposed into $k$ disjoint sets ($V_1$, $V_2$,...,$V_k$) such that no two graph vertices within the same set are adjacent) such that every pair of graph vertices in the $k$ sets are adjacent (\Cref{fig:Delta_graph}).
\end{definition}
In terms of naming, let $K_{x,y}$ be a complete bi-partite graph, and $K_{x,y,z}$ be a complete three-partite graph; $x$, $y$ and $z$ correspond to the number of vertices included into each vertices subset. For instance, the $K_{3,2}$ depicted in \Cref{fig:2Y_graph} presents a couple of vertices subsets $V_1=\left\lbrace v_2,v_3,v_4 \right\rbrace $ and $V_2 =\left\lbrace v_1,v_5 \right\rbrace $ so that $x=3$ and $y=2$. The $K_{1,1,1}$ depicted in \Cref{fig:Delta_graph} presents a triple of vertices subsets $V_1=\left\lbrace v_1\right\rbrace $, $V_2 =\left\lbrace v_2 \right\rbrace $ and $V_3 =\left\lbrace v_3 \right\rbrace $ composed by one node respectively, so that $x=1$, $y=1$ and $z=1$.

\subsection{Graph Laplacian}
The topologies can be represented through the graph Laplacian, or Kirchhoff matrix. On this regard, recalling \cite{chung1996lectures,chung1996combinatorial}, a few mathematical definitions are provided. For a given graph $\mathcal{G}\left(V_\mathcal{G},E_\mathcal{G}\right) $ with $n$ vertices $V_\mathcal{G} = (v_1 , v_2,..., v_n)$, $m$ edges $E_\mathcal{G} = (e_1 , e_2,..., e_m)$, and let $d_v$ denotes the degree of the vertex $v$, that is, the number of edges attached to each vertex, so, the graph Laplacian $L_{n\times n}$ is defined element-wise as follows:
	\begin{align}
	\mathbf{L}_{\mathcal{G}}(i,j) & = 
	\begin{cases}
		d_v  & \text{if } i=j\\
		-1 & \text{if } i \neq j \text{ and } v_i \text{ is adjacent to } v_j \\
		0 & \text{otherwise}
	\end{cases}
\end{align}
Analogously, it can be also expressed as follows:
\begin{equation}
	\mathbf{L}_\mathcal{G} = \mathbf{B}_\mathcal{G}^\top \mathbf{B}_\mathcal{G}  \quad\text{or}\quad \mathbf{L}_\mathcal{G} = \mathbf{D}_\mathcal{G} - \mathbf{A}_\mathcal{G},
\end{equation}
where $\mathbf{B}_\mathcal{G}$ and $\mathbf{A}_\mathcal{G}$ represent the \textit{incidence} and \textit{adjacency} matrix respectively of $\mathcal{G}$. $\mathbf{D}_\mathcal{G}$ is the so-called \textit{degree} diagonal matrix. In case of a weighted graph $\mathcal{G}\left(V_\mathcal{G},E_\mathcal{G},w_\mathcal{G}\right) $, it becomes:
\begin{equation}
	\mathbf{L}_\mathcal{G} = \mathbf{B}_\mathcal{G}^\top \mathbf{W}_\mathcal{G}  \mathbf{B}_\mathcal{G},
\end{equation}
where the $\mathbf{W}_{\mathcal{G},m\times m}$ is a diagonal matrix including weight $w_{ij}$ at the element $i=j$. As described in \cite{spielman2012spectral,chung1996lectures}, the knowledge of the matrix $\mathbf{L}_\mathcal{G}$ allows to assess the spectral graph properties. 
The eigenvectors-eigenvalues matrices ($\mathbf{P}_\mathcal{G} $ and $\mathbf{M}_\mathcal{G} $) obtained by diagonalizing the Laplacian matrix, contain the physical normal modes, also called \textit{vibrational} modes, at the nodes, which correspond to the nodal DOFs of the graph topology. Being linked to the nodes, these can be further analysed to identify possible coupling-decoupling among the graph vertices. From \cite{strang2019linear}, a graph Laplacian matrix $L_{n\times n}$ is square and symmetric and \textit{positive semidefinite}: it has $n-1$ eigenvalues and one zero eigenvalue. Furthermore, the complete bi-partite graphs eigenvalues patterns are established by the following definition:
\begin{definition}[Complete bi-partite graph eigenvalues]
	A complete bipartite graph $K_{x,y}$ presents eigenvalues $x+y$, $y$, $x$ and $0$; with multiplicity $1$, $x-1$, $y-1$ and $1$ respectively. For instance: let $K_{3,2}$ be the complete bi-partite graph, the corresponding eigenvalues are, then: $\lambda_{K_{3,2}}=\left[5,2,2,3,0\right] $. Then $\mathbf{M}_\mathcal{G} = \mathrm{diag}\left( \lambda_{K_{3,2}} \right)  $.
\end{definition}
The eigenvector matrix $\mathbf{P}_\mathcal{G}$ can be used to transit from nodal variables, $\mathbf{x}_{v,\mathcal{G}}$, to  
nodal normal modes, $\mathbf{x}_{\mathrm{dec},\mathcal{G}}$, as follows:
\begin{equation} \label{eq:nodes_modes}
	\mathbf{x}_{v,\mathcal{G}} =  \mathbf{P}_{\mathcal{G}}  \mathbf{x}_{\mathrm{dec},\mathcal{G}} \quad\text{;}\quad \mathbf{x}_{\mathrm{dec},\mathcal{G}} =  \mathbf{P}_{\mathcal{G}}^{-1}  \mathbf{x}_{v,\mathcal{G}},
\end{equation}
where: $\mathbf{x}_{\mathcal{G}} = \begin{bmatrix}
	x_{1,\mathcal{G}} & x_{2,\mathcal{G}} & \cdots & x_{n,\mathcal{G}} 
\end{bmatrix}^\top $.\\
\textbf{Remark}: \textit{The graph Laplacian eigenvectors, $\mathbf{P}_\mathcal{G}$, do provide the all possible graph nodal DOFs. Revealing linear decoupled patterns among the nodes, it is possible to recognize eventual affinities between each other, which can be useful to identify ports affinities within an electrical sub-network}.\\
With respect to the edges, for an unweighted graph, let $\textbf{x}_{e,\mathcal{G}}$ be the edges variables of the graph, then, the relationship with the decoupled set of nodes variables is the following:
\begin{equation} \label{eq:edges_variables_graph}
	 \mathbf{x}_{e,\mathcal{G}} = \mathbf{B}_\mathcal{G}  \mathbf{P}_\mathcal{G}  \mathbf{x}_{\mathrm{dec},\mathcal{G}} = \mathbf{B}_\mathcal{G}\mathbf{x}_{v,\mathcal{G}}.
\end{equation}
where: $\mathbf{x}_{e,\mathcal{G}} = \begin{bmatrix}
	x_{1,\mathcal{G}} & x_{2,\mathcal{G}} & \cdots & x_{m,\mathcal{G}} 
\end{bmatrix}^\top $.\\
\textbf{Remark}: \textit{To be noticed that the edges variables can be reconstructed by the nodal normal modes superposition}.

\subsection{Electrical Network}
To analyse the characteristic nodal normal modes of a certain electrical topology $\mathcal{G}$, momentarily, it can be possible to refer to a passive sub-network as depicted in \Cref{fig:ext_active}. The terminals represent the points of connection with the external sources; while the arm conductances (inductances are not considered momentarily), establish the electrical interconnectivity through the nodes. From now on, the analysis is conducted by considering, for each arm constituting the topology $\mathcal{G}$, the same conductance $G_\mathrm{a}$, then:
\begin{equation} \label{eq:homogeneous_edges}
	\mathbf{W}_\mathcal{G} = G_\mathrm{a} \mathbf{I}_{m} \quad;\quad \mathbf{I}_{m} = \mathrm{diag}\left(1,1,\dots,1 \right) 
\end{equation}
\subsubsection{Terminals DOFs}
Let $\mathcal{G}$ be a sub-network graph, then the voltage and current node-edge relationships are expressed below:
\begin{equation} \label{eq:nodes}
	\mathbf{i}_{v,\mathcal{G}}  =  \mathbf{B}_\mathcal{G}^\top  \mathbf{i}_{e,\mathcal{G}} \quad\mathrm{;}\quad \mathbf{u}_{e,\mathcal{G}} = \mathbf{B}_\mathcal{G}  \mathbf{u}_{v,\mathcal{G}},
\end{equation}
where $\mathbf{x}_{v,\mathcal{G}}$ and $\mathbf{x}_{e,\mathcal{G}}$ represent the variables vectors at the terminals and across the arms, respectively. 
Furthermore, the nodal voltage-current relationship is provided by the Laplacian matrix:
\begin{equation}\label{eq:nodal_relation}
	\mathbf{L}_\mathcal{G}\mathbf{u}_{v,\mathcal{G}} = \mathbf{i}_{v,\mathcal{G}} \quad\text{where}\quad \mathbf{L}_\mathcal{G} = \mathbf{B}_\mathcal{G}^\top G_\mathrm{a} \mathbf{I}_{m} \mathbf{B}_\mathcal{G}.
\end{equation}
\Cref{eq:nodal_relation} represents the nodal electrical charge balancing equation. By diagonalizing the Laplacian matrix and referring to \Cref{eq:nodes_modes}, a decoupled variable patterns are achieved:
\begin{equation}\label{eq:pure_resistive_network_1}
	\mathbf{M}_{\mathcal{G}_r}\mathbf{u}_{\mathrm{dec},\mathcal{G}} = \mathbf{i}_{\mathrm{dec},\mathcal{G}} \quad\mathrm{;}\quad \mathbf{M}_{\mathcal{G}_r,n\times n} = G_\mathrm{a}\mathrm{diag}\left(\lambda_1,\dots ,\lambda_n \right).
\end{equation}
Where $\mathbf{u}_{\mathrm{dec},\mathcal{G}}$ and $\mathbf{i}_{\mathrm{dec},\mathcal{G}}$ indicate the vectors of decoupled voltage and current at the vertices of $\mathcal{G}$. $\mathbf{M}_{\mathcal{G}_r}$ represents diagonal eigenvalues matrix considering infinite the external system conductances values. For a predominantly inductive arms, inductance $L_\mathrm{a}$, the decoupled system of equations can be rewritten as follows:
\begin{equation}\label{eq:a}
	\mathbf{M}_{\mathcal{G}_l}\mathbf{u}_{\mathrm{dec},\mathcal{G}} = \dfrac{\mathrm{d}}{\mathrm{d}t }\mathbf{i}_{\mathrm{dec},\mathcal{G}} \quad\mathrm{;}\quad 	\mathbf{M}_{\mathcal{G}_l,n\times n} = L_\mathrm{a}^{-1}\mathrm{diag}\left(\lambda_1,\dots ,\lambda_n \right).
\end{equation}
The constants values $\left( \lambda_1 \leq \lambda_2 \leq \dots \leq \lambda_n\right)  \in \mathbb{R}$ are the characteristic graph Laplacian matrix eigenvalues of $\mathcal{G}$.

\subsubsection{External Systems Conductances}\label{sssec:external_parameters}
The additional external systems conductances leads to a different overall Laplacian matrix of $\mathcal{G}$. In fact, let $G_\mathrm{a}$ be the internal converter arm nodal conductance, and $G_\mathrm{ext}$ be the external system nodal conductance; the overall terminal conductance $G_\mathrm{s}$ is treated as the series connected conductances as depicted in \Cref{eq:series_cond}. However, when the external conductance $G_\mathrm{ext}\gg G_\mathrm{int}$, the overall conductance $G_\mathrm{s} \backsimeq G_\mathrm{a}$ and vice-versa:  
\begin{equation}\label{eq:series_cond}
G_\mathrm{s} = \dfrac{G_\mathrm{a} G_\mathrm{ext}}{G_\mathrm{a} + G_\mathrm{ext}} \quad\text{;}\quad   \lim_{G_\mathrm{ext}\to\infty} G_\mathrm{s} = G_\mathrm{a}.
\end{equation}
In a more complete form, the \Cref{eq:nodal_relation} can be rewritten as follows: 
\begin{equation}\label{eq:complete_laplacian}
	\underbrace{\left(\dfrac{\mathbf{L}_{\mathcal{G}_{r,\mathrm{int}}} \mathbf{G}_{\mathcal{G}_{\mathrm{ext}}}}{\mathbf{L}_{\mathcal{G}_{r,\mathrm{int}}} + \mathbf{G}_{\mathcal{G}_{\mathrm{ext}}}} \right)}_{\mathbf{L}_{\mathcal{G}_{r,\mathrm{tot}}}} \mathbf{u}_{v,\mathcal{G}} = \mathbf{i}_{v,\mathcal{G}},
\end{equation} 
where $\mathbf{G}_{\mathcal{G}_{\mathrm{ext}}}$ represents the diagonal matrix of the external conductances at the nodes $V_\mathcal{G} = (v_1 , v_2,..., v_n)$:
\begin{equation}
	\mathbf{G}_{\mathcal{G}_{\mathrm{ext}}} = \begin{bmatrix}
		G_{\mathrm{ext},1} & G_{\mathrm{ext},2} & \dots & G_{\mathrm{ext},n}
	\end{bmatrix}.
\end{equation}
Then, based on $\mathbf{L}_{\mathcal{G}_{r,\mathrm{tot}}}$, the eigenvectors-eigenvalues matrices $\mathbf{P}_{\mathcal{G}_{r,\mathrm{tot}}}$ and $\mathbf{M}_{\mathcal{G}_{r,\mathrm{tot}}}$ of the topology $\mathcal{G}$ are revisited.

\subsubsection{Nodal normal modes from external-internal sources}
\begin{figure}[!h]
	\centering
	\subfloat[]
	{
	\includegraphics[]{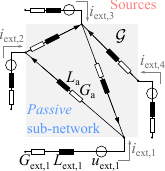}
		\label{fig:ext_active}
	}
	\hspace*{-1em}
	\subfloat[]
	{
	\includegraphics[]{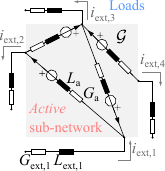}
		\label{fig:int_active}
	}
	\hspace*{-1em}
	\subfloat[]
	{
	\includegraphics[]{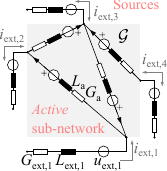}
		\label{fig:ext_int}
	}
	\caption{Nodal normal modes forced by: external sources in (a), internal sources in (b) and by both, external and internal sources in (c).}
	\label{fig:network_configurations}
\end{figure}

From \Cref{eq:nodes_modes} and \Cref{eq:edges_variables_graph}, it is important to specify that nodal, and edges, voltage-current variables of a sub-network $\mathcal{G}$ can be forced either by the external nodal sources or internal edge sources. Consequently, the expected nodal normal modes of the topology $\mathcal{G}$, generated by the internal and external sources, interact to each other to establish a nodal current flow. \Cref{fig:ext_active} is a clear case of unidirectional nodal normal modes generated by external nodal sources. Vice-versa, in \Cref{fig:int_active}, the unidirectional current nodal normal modes are provoked by internal edges sources. The enabled bidirectional nodal normal modes current flow is illustrated in \Cref{fig:ext_int}.

\begin{figure}[!h]
	\centering
	\subfloat[]
	{
		\includegraphics[]{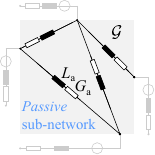}
				\label{fig:no_loop}
	}
	\hspace*{-1em}
	\subfloat[]
	{
		\includegraphics[]{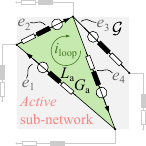}
				\label{fig:loop_1}
	}
	\hspace*{-1em}
	\subfloat[]
	{
		\includegraphics[]{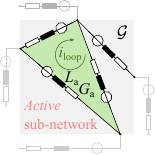}
				\label{fig:loop_2}
	}
	\caption{In (a) a case of unavailable internal current loop. Available internal current loop without or with external nodal sources applied in (b) and (c) respectively.}
	\label{fig:internal_loops_network}
\end{figure}

\subsubsection{Internal circulating current DOFs}
A cell-based converter topology, contrarily to the circuit \Cref{fig:no_loop}, due to the presence of the arm controllable voltage sources, can introduce internal circulating current loops. Furthermore, the internal current loops are totally decoupled from the nodal normal modes variables of the circuit, \Cref{fig:loop_1} and \Cref{fig:loop_2}.\\
Let $k$ be the number of the available internal current loops of a sub-network converter topology $\mathcal{G}$ (referring to \Cref{fig:internal_loops_network}, $k=1$) with a given current orientation, it is possible to define a matrix $\mathcal{N}_{\mathcal{G},k\times m}$ which is defined element-wise as: $\mathcal{N}_{\mathcal{G}} (i,j) = \mathrm{sign}(i_e)$.
Referring to \Cref{fig:loop_1}: $\mathcal{N}_{\mathcal{G},1\times 4} = \begin{bmatrix}
1 & 1 & 0 & 1 
\end{bmatrix}$.\\
Then, the internal current loops variables $\mathbf{i}_{\mathrm{loop},\mathcal{G}}$ can be calculated as follows:
\begin{equation} \label{eq:int_dec_description_1}
	G_\mathrm{a} \mathcal{N}_{\mathcal{G},k\times m} \mathbf{u}_{e,\mathcal{G}} = \mathbf{i}_{\mathrm{loop},\mathcal{G}}.
\end{equation}
Therefore, aiming to achieve the effective independent internal current loops DOFs, $\mathbf{i}_{\mathcal{G},\mathrm{dec,loop}}$; the $\mathrm{rank}(\mathcal{N}_{\mathcal{G},k\times m}^\top)$ indicates the number of the independent loops and, an orthonormal basis $\mathcal{B}_{\mathrm{loop},\mathcal{G}}$ can be calculated to achieve internal current loops DOFs relationship with the arm current variables $\mathbf{i}_{e,\mathcal{G}}$:
\begin{equation} \label{eq:int_dec_description}
	\mathbf{i}_{\mathcal{G},\mathrm{dec,loop}} = \mathcal{B}_{\mathrm{loop},\mathcal{G}}^\top \mathbf{i}_{e,\mathcal{G}} \quad\text{and}\quad \mathbf{u}_{\mathcal{G},\mathrm{dec,loop}} = \mathcal{B}_{\mathrm{loop},\mathcal{G}}^\top \mathbf{u}_{e,\mathcal{G}}.
\end{equation}
Furthermore, recalling \Cref{eq:edges_variables_graph}, by fully exploiting the topology DOFs, the arm current variables of a topology $\mathcal{G}$ are defined as the following superposition:
\begin{equation} \label{eq:edges_variables_complete}
	\mathbf{i}_{e,\mathcal{G}}  =  \underbrace{\mathbf{B}_\mathcal{G}  \mathbf{P}_\mathcal{G}  \mathbf{M}_\mathcal{G}^{-1} \mathbf{i}_{\mathrm{dec},\mathcal{G}}}_{\text{from nodal DOFs}} + \underbrace{\mathcal{B}_{\mathrm{loop},\mathcal{G}} \mathbf{i}_{\mathcal{G},\mathrm{dec,loop}}}_\text{from internal DOFs}.
\end{equation}
For a most general scenario, where both the external sources and internal converter arm sources, which are also interconnected to enable the internal circulating current loops, the full decoupled set of equations can be summarized as follows:
\begin{equation}\label{eq:resistive_case}
	\begin{cases}
		& \underbrace{\mathbf{P}_{\mathcal{G}_{r,\mathrm{tot}}}^{-1} \mathbf{u}_{v,\mathrm{ext}}}_{\text{ext. source}} - \underbrace{\mathbf{P}_{\mathcal{G}_{r,\mathrm{tot}}}^{-1} \mathbf{u}_{v,\mathrm{int}}}_{\text{int. sources}} =\mathbf{M}_{\mathcal{G}_{r,\mathrm{tot}}}^{-1} \mathbf{P}_{\mathcal{G}_{r,\mathrm{tot}}}^{-1} \mathbf{i}_{v,\mathcal{G}} \\[5pt]
		& - \mathbf{u}_{\mathcal{G},\mathrm{dec,loop}}G_\mathrm{a} =  \mathbf{i}_{\mathcal{G},\mathrm{dec,loop}}.
	\end{cases}
\end{equation}
For a predominantly inductive parameters of the circuit, the system is rewritten as follows:
\begin{equation}\label{eq:inductive_case}
	\begin{cases}
		& \underbrace{\mathbf{P}_{\mathcal{G}_{r,\mathrm{tot}}}^{-1} \mathbf{u}_{v,\mathrm{ext}}}_{\text{ext. source}} - \underbrace{\mathbf{P}_{\mathcal{G}_{r,\mathrm{tot}}}^{-1} \mathbf{u}_{v,\mathrm{int}}}_{\text{int. sources}} =\mathbf{M}_{\mathcal{G}_{r,\mathrm{tot}}}^{-1} \dfrac{\mathrm{d}}{\mathrm{d}t} \left( \mathbf{P}_{\mathcal{G}_{r,\mathrm{tot}}}^{-1} \mathbf{i}_{v,\mathcal{G}}\right)  \\[5pt]
		& - \mathbf{u}_{\mathcal{G},\mathrm{dec,loop}}L_\mathrm{a}^{-1} =  \dfrac{\mathrm{d}}{\mathrm{d}t} \mathbf{i}_{\mathcal{G},\mathrm{dec,loop}}.
	\end{cases}
\end{equation}
\Cref{eq:resistive_case} and \Cref{eq:inductive_case} can be eventually added to each other .\\
\textbf{Remark}: \textit{Each DOF equation refers to an equivalent single-phase circuit. Depending on the specific application, each DOF can be treated either as DC or AC domain}.

\subsubsection{Power Degrees of Freedom}
The idea behind the instantaneous power theory presented in \cite{akagi2017instantaneous}, referred to the $\mathcal{Y}$ based topology, is, through this paragraph, further elaborated and generalized to characterize the power DOFs at both the terminals and arm converters levels of different topologies. In fact, if $\mathcal{G}$ is a converter sub-network with $n$ terminals and $m$ arms, the corresponding nodal power $\mathbf{p}_{v,\mathcal{G}}$ and arm power $\mathbf{p}_{e,\mathcal{G}}$ are defined as:
\begin{equation}
	\begin{cases}
		& \mathbf{p}_{v,\mathcal{G}}= \begin{bmatrix}
			p_{1,\mathcal{G}} & p_{2,\mathcal{G}} & \cdots & p_{n,\mathcal{G}} 
			\end{bmatrix}^\top\\[5pt]
		&\mathbf{p}_{e,\mathcal{G}}= \begin{bmatrix}
			p_{1,\mathcal{G}} & p_{2,\mathcal{G}} & \cdots & p_{m,\mathcal{G}} 
			\end{bmatrix}^\top.
	\end{cases}
\end{equation}
Referring to the voltage-current DOFs, $\mathbf{u}_{\mathrm{dec},\mathcal{G}}$ and $\mathbf{i}_{\mathrm{dec},\mathcal{G}}$, the instantaneous power expressions are formulated as follows:
\begin{equation}\label{eq:node_power}
	\mathbf{p}_{v,\mathcal{G}}(t) = \mathbf{u}_{v,\mathcal{G}}(t) \odot \textbf{i}_{v,\mathcal{G}}(t) = \mathbf{\Gamma}_{n \times h}'\left[ \mathbf{u}_{\mathrm{dec},\mathcal{G}}(t) \otimes \textbf{i}_{\mathrm{dec},\mathcal{G}}(t)\right],
\end{equation}
\begin{equation} \label{eq:edge_power}
	\mathbf{p}_{e,\mathcal{G}}(t)  =  \mathbf{u}_{e,\mathcal{G}}(t) \odot \textbf{i}_{e,\mathcal{G}}(t) = \mathbf{\Gamma}_{m \times h}'' \underbrace{\left[ \mathbf{u}_{\mathrm{dec},\mathcal{G}}(t) \otimes \textbf{i}_{\mathrm{dec},\mathcal{G}}(t)\right]}_{\mathbf{p}_\mathcal{G}^\mathrm{dec}(t)} .   
\end{equation}
Where the operators $\odot$ and $\otimes$ represent the element-wise product and the tensor product respectively. Such that, for instance, for a given voltage variables vector $\mathbf{u}(t)$ and a given current variables vector $\mathbf{i}(t)$: 
\begin{equation} \label{eq:ex1}
	\mathbf{u}(t) = \begin{bmatrix}
		u_1 & u_2
	\end{bmatrix}^\top \quad;\quad\mathbf{i}(t) = \begin{bmatrix}
	i_\mathrm{a} & i_\mathrm{b}
	\end{bmatrix}^\top;
\end{equation}
the element-wise product operator $\odot$ performs as follows:
\begin{equation} \label{eq:ex2}
	\mathbf{p}(t)_\odot = \begin{bmatrix}
		u_1i_\mathrm{a} & u_2i_\mathrm{b}
	\end{bmatrix}^\top,
\end{equation}
while the tensor product operator $\otimes$ performs as follows:
\begin{equation} \label{eq:ex3}
	\mathbf{p}(t)_\otimes = \begin{bmatrix}
		u_1i_\mathrm{a} & u_1i_\mathrm{b} & u_2i_\mathrm{a} & u_2i_\mathrm{b}
	\end{bmatrix}^\top,
\end{equation}
which is collecting the all possible product combinations between the elements of the two vectors.\\
Matrices $\mathbf{\Gamma}_{n \times h}'$ and $\mathbf{\Gamma}_{m \times h}''$, with elements $\mathbf{\Gamma}_{n \times h}'(i,j) \in \mathbb{R} $ and $\mathbf{\Gamma}_{m \times h}''(i,j) \in \mathbb{R} $, express the nodal power variables and the edges power variables respectively as a linear combination of the power components collected in the vector $\mathbf{p}_\mathcal{G}^\mathrm{dec}(t)$.\\
Then the node-edge instantaneous power expressions, of a certain topology $\mathcal{G}$, can be formulated with the corresponding current-voltage DOFs components: $\mathbf{i}_{\mathrm{dec},\mathcal{G}}(t)$ and $\mathbf{u}_{\mathrm{dec},\mathcal{G}}(t)$.\\
\textbf{Remark}: \textit{Each row of $\mathbf{p}_{v,\mathcal{G}}$ and $\mathbf{p}_{e,\mathcal{G}}$ describes a single-phase power. The nature of the decoupled patterns variables marks the power terms frequencies. In general, oscillatory and constant power terms are expected}.\\
Additionally, by analysing matrices $\mathbf{\Gamma}_{n \times h}'$ and $\mathbf{\Gamma}_{m \times h}''$, it can be possible to define the number of linear independent powers $N_\mathrm{dec}'$ and $N_\mathrm{dec}''$ and find a basis $\mathcal{B}_{\mathrm{p},v}$ and $\mathcal{B}_{\mathrm{p},e}$ respectively:
\begin{equation} \label{eq:nodes1}
	\mathrm{rank}\left(\mathbf{\Gamma}_{n \times h}' \right)  =  N_\mathrm{dec}' \quad\mathrm{;}\quad \mathcal{B}_{\mathrm{p},v} = \left\lbrace \textbf{t}_1,\dotsc, \mathbf{t}_{N_\mathrm{dec}'} \right\rbrace ,
\end{equation}
\begin{equation} \label{eq:nodes2}
	\mathrm{rank}\left(\mathbf{\Gamma}_{m \times h}'' \right)  =  N_\mathrm{dec}'' \quad\mathrm{;}\quad \mathcal{B}_{\mathrm{p},e} = \left\lbrace \mathbf{t}_1,\dotsc, \mathbf{t}_{N_\mathrm{dec}''} \right\rbrace .
\end{equation}
Finally, for a graph $\mathcal{G}$, the decoupled instantaneous power patterns at nodes and at the edges, $\mathbf{p}_{v,\mathcal{G}}^{\mathrm{new}}(t)$ and $\mathbf{p}_{e,\mathcal{G}}^{\mathrm{new}}(t)$, respectively, are summarized below:
\begin{equation} \label{eq:decoupled_power_basis}
	\begin{cases}
		&	\mathbf{p}_{v,\mathcal{G}}^{\mathrm{new}}(t) = \mathcal{B}_{\mathrm{p},v} \mathbf{p}_{v,\mathcal{G}}(t)\\
		&	\mathbf{p}_{e,\mathcal{G}}^{\mathrm{new}}(t) = \mathcal{B}_{\mathrm{p},e} \mathbf{p}_{e,\mathcal{G}}(t).
	\end{cases}
\end{equation}
As for the power, the terminal-arms energies of the topology $\mathcal{G}$ , $\mathbf{e}_{v,\mathcal{G}}(t)$ and $\mathbf{e}_{e,\mathcal{G}}(t)$, can be then formulated by employing the characteristic voltage-current variables DOFs; in fact:
\begin{equation} \label{eq:nodes3}
	\begin{cases} 
		&	\mathbf{e}_{v,\mathcal{G}}(t) = \textbf{E}_{v,\mathcal{G}}^i + \displaystyle \int_{t_0}^\tau \mathbf{p}_{v,\mathcal{G}}(t) \mathrm{d}t \\[10pt]
		&	\mathbf{e}_{e,\mathcal{G}}(t) = \textbf{E}_{e,\mathcal{G}}^i + \displaystyle \int_{t_0}^\tau \mathbf{p}_{e,\mathcal{G}}(t) \mathrm{d}t ,
	\end{cases}
\end{equation}
where $\mathbf{E}_{v,\mathcal{G}}^i$, $\mathbf{E}_{e,\mathcal{G}}^i$ are the initial energy values at the nodes and at the edges respectively.

\section{Converter Topologies DoFs Analysis}\label{sec:application}
By referring to the complete $k$-partite graph topologies depicted in \Cref{fig:common_topologies}, the aim of the section is to guide toward the systematic converter topologies DOFs assessment. The methodology is summarized by the following points:  
\begin{itemize}
	\item Find the Laplacian matrix of the topology and calculate the corresponding eigenvectors-eigenvalues matrices.
	\item Establish the node-edge voltage-current relationship referring to the topology normal modes.
	\item Identify, and include, eventual independent nested current loops to fully describe the edges current variables.
	\item Formulate the instantaneous edge-node power patterns based on voltage-current topology DOFs. 
\end{itemize}
The eigenvectors-eigenvalues analysis is settled considering external conductances parameters times larger than the arm conductances and negligible external inductances. Refer to \Cref{eq:complete_laplacian} for a more generalized system analysis.
\vspace{-2mm}
\subsection{$\mathcal{I}$ Topology}
The complete bi-partite graph circuit topology $K_{1,1}$, depicted in \Cref{fig:I_graph}, can be linked to the cell-based converter proposed in \cite{baker1975electric}. The one-arm converter topology can be described by the incidence matrix and the graph Laplacian: 
\begin{equation}
	\mathbf{B}_\mathcal{I} = \left(\begin{matrix*}[r]
		-1 & 1\\
	\end{matrix*}\right) \quad\mathrm{;}\quad \mathbf{L}_\mathcal{I}= \left(\begin{matrix*}[r]
	1 & -1\\
	-1 & 1
	\end{matrix*}\right).
\end{equation}
At the vertices, there exists the following relationship:
\begin{equation}
	\mathbf{L}_\mathcal{I}\mathbf{u}_{v,\mathcal{I}} = \mathbf{i}_{v,\mathcal{I}} \quad\text{where}\quad \mathbf{x}_{v,\mathcal{I}} = \begin{bmatrix}
		x_{v_1} & x_{v_2}
	\end{bmatrix}^\top.
\end{equation}
Applying \Cref{eq:pure_resistive_network_1} the nodal decoupled variables are achieved:
\begin{equation} \label{eq:decoupled_I}
	\mathbf{M}_\mathcal{I} \mathbf{u}_{\mathrm{dec},\mathcal{I}} = \mathbf{i}_{\mathrm{dec},\mathcal{I}} \quad\mathrm{;}\quad \mathbf{M}_\mathcal{I} = G_\mathrm{a}\mathrm{diag}\left( 0, 4\right) .
\end{equation}
The eigenvector matrix $\mathbf{P}_\mathcal{I}$ performs transformation between the normal nodes variables and the decoupled set of variables:
\begin{equation}
	\mathbf{u}_{\mathrm{dec},\mathcal{I}} = \mathbf{P}_{\mathcal{I}}^{-1} \mathbf{u}_{v,\mathcal{I}} \quad\mathrm{;}\quad  \textbf{i}_{\mathrm{dec},\mathcal{I}} = \mathbf{P}_{\mathcal{I}}^{-1} \mathbf{i}_{v,\mathcal{I}}
\end{equation}
where:
\begin{equation}
	\mathbf{P}_\mathcal{I}^{-1} = \dfrac{1}{2}\left(\begin{matrix*}[r]
		1 & 1\\
		-1 & 1
	\end{matrix*}\right)  \quad\mathrm{;}\quad  \mathbf{x}_{\mathrm{dec},\mathcal{I}} = \left(\begin{matrix*}[r]
	x_{\mathrm{dec},\mathcal{I}}^{\lambda_0}\\
	x_{\mathrm{dec},\mathcal{I}}^{\lambda_\alpha}
	\end{matrix*}\right).
\end{equation}
and $\mathbf{x}_{\mathrm{dec},\mathcal{I}}$ summarizes the available modes referred to the eigenvalues $\lambda_0$ and $\lambda_\alpha$. Since the graph does not count any internal loop, the arm converter $e_1$ variables are below:
\begin{equation}
	u_{e_{1},\mathcal{I}} = \mathbf{B}_\mathcal{I} \mathbf{P}_{I} \mathbf{u}_{\mathrm{dec},\mathcal{I}} \quad\mathrm{;}\quad i_{e_{1},\mathcal{I}} = \mathbf{B}_\mathcal{I} \mathbf{P}_{\mathcal{I}} \mathbf{M}_I^{-1} \textbf{i}_{\mathrm{dec},\mathcal{\mathcal{I}}},
\end{equation}
while the power:
\begin{equation}
	p_{e_{1},\mathcal{I}} = u_{e_{1},\mathcal{I}}i_{e_{1},\mathcal{I}} = 2 u_{\mathrm{dec},\mathcal{I}}^{\lambda_\alpha}  i_{\mathrm{dec},\mathcal{I}}^{\lambda_\alpha} = (u_{v_{2}} - u_{v_{1}})i_{e_{1}}.
\end{equation}

\subsection{$\mathcal{V}$ Topology}
The complete bi-partite graph circuit topology $K_{1,2}$, depicted in \Cref{fig:V_graph}, also called \textit{open-$\Delta$}, is inspiring alternative converter arrangements \cite{kang1999open,xiang2020analysis}. The circuit topology is described by the incidence matrix and the graph Laplacian:
\begin{equation}
	\mathbf{B}_{\mathcal{V}} = \left(\begin{matrix*}[r]
		-1 & 1 & 0\\
		-1 & 0 & 1
	\end{matrix*}\right) \quad\mathrm{;}\quad 	\mathbf{L}_{\mathcal{V}}= \left(\begin{matrix*}[r]
	2 & -1 & -1\\
	-1 & 1 & 0\\
	-1 & 0 & 1
	\end{matrix*}\right).
\end{equation}
At the vertices, there exists the following relationship:
\begin{equation}
	\mathbf{L}_{\mathcal{V}}\mathbf{u}_{v,\mathcal{V}} = \mathbf{i}_{v,\mathcal{V}} \quad\text{where}\quad \mathbf{x}_{v,\mathcal{V}} = \begin{bmatrix}
		x_{v_1} & x_{v_2} & x_{v_3}
	\end{bmatrix}^\top.
\end{equation}
Applying \Cref{eq:pure_resistive_network_1} the nodal decoupled variables are achieved:
\begin{equation} \label{eq:decoupled_I}
	\mathbf{M}_{\mathcal{V}} \mathbf{u}_{\mathrm{dec},\mathcal{V}} = \mathbf{i}_{\mathrm{dec},\mathcal{V}} \quad\mathrm{;}\quad \mathbf{M}_{\mathcal{V}} = G_\mathrm{a}\mathrm{diag}\left( 0, 1,3\right) .
\end{equation}
And:
\begin{equation}
	\mathbf{P}_{\mathcal{V}}^{-1} = \dfrac{1}{3}\left(\begin{matrix*}[r]
		1 & 1 & 1\\
		0 & -3/2 & 3/2\\
		-1 & 1/2 & 1/2
	\end{matrix*}\right) \quad\mathrm{;}\quad  \mathbf{x}_{\mathrm{dec},\mathcal{V}} = \left(\begin{matrix*}[r]
	x_{\mathrm{dec},\mathcal{V}}^{\lambda_0}\\
	x_{\mathrm{dec},\mathcal{V}}^{\lambda_\alpha}\\
	x_{\mathrm{dec},\mathcal{V}}^{\lambda_\beta}
	\end{matrix*}\right)
\end{equation}
perform the following transformation:
\begin{equation}
	\mathbf{u}_{\mathrm{dec},\mathcal{V}} = \mathbf{P}_{\mathcal{V}}^{-1} \mathbf{u}_{v,\mathcal{V}} \quad\mathrm{;}\quad  \mathbf{i}_{\mathrm{dec},\mathcal{V}} = \mathbf{P}_{\mathcal{V}}^{-1} \mathbf{i}_{v,\mathcal{V}}.
\end{equation}
From \Cref{eq:edges_variables_graph}, since the graph does not count any internal loop, the arm converters variables $\mathbf{x}_{e,\mathcal{V}} = \begin{bmatrix}
	x_{e_1} & x_{e_2}
\end{bmatrix}^\top$ are below:
\begin{equation}
	\mathbf{u}_{e,\mathcal{V}} =  \left(\begin{matrix*}[r]
		0 & -1 & 3\\
		0 & 1 & 3
	\end{matrix*}\right) \mathbf{u}_{\mathrm{dec},\mathcal{\mathcal{V}}} \quad\mathrm{;}\quad \mathbf{i}_{e,\mathcal{V}} =  \left(\begin{matrix*}[r]
		0 & -1 & 1\\
		0 & 1 & 1
	\end{matrix*}\right) \mathbf{i}_{\mathrm{dec},\mathcal{V}}.
\end{equation}
From now on, for a more concise notation, mixed power term $i_{\mathrm{dec},\mathcal{G}}^{\lambda_x}u_{\mathrm{dec},\mathcal{G}}^{\lambda_y}$, will be replaced as $p_{\mathrm{d},\mathcal{G}}^{xy}$. Recalling \Cref{eq:edge_power}, the instantaneous edges powers are summarized below:
\begin{equation} \label{eq:edge_V_power}
	\mathbf{p}_{e,\mathcal{V}} = \left(\begin{matrix*}[r]
		1 & -3 & -1 & 3\\
		1 & 3 & 1 & 3
	\end{matrix*}\right) \mathbf{p}_{e,\mathcal{V}}^\mathrm{dec}
\end{equation}
where:
\begin{equation}
	\mathbf{p}_{e,\mathcal{V}}^\mathrm{dec} = \begin{bmatrix}
		p_{\mathrm{d},\mathcal{V}}^{\alpha\alpha} & p_{\mathrm{d},\mathcal{V}}^{\alpha\beta} & p_{\mathrm{d},\mathcal{V}}^{\beta\alpha} & p_{\mathrm{d},\mathcal{V}}^{\beta\beta}
	\end{bmatrix}^\top,
\end{equation}
which is not affected by the decoupled mode $\lambda_0$. Based on \Cref{eq:node_power}, the nodal powers are defined below:
\begin{equation} \label{eq:V_power}
	\mathbf{p}_{v,\mathcal{V}} = 
	\begin{pmatrix*}[r]
		0 & 0 & 0 & 4 & 1 & -2 & 0 & -2 & 0\\
		1 & -1 & -1 & 1 & 1 & 1 & -1 & 1 & -1\\
		1 & 1 & 1 & 1 & 1 & 1 & 1 & 1 & 1
	\end{pmatrix*}
	\mathbf{p}_{v,\mathcal{V}}^\mathrm{dec} 
\end{equation}
where:
\begin{equation}
	\mathbf{p}_{v,\mathcal{V}}^\mathrm{dec} = \begin{bmatrix}
		p_{\mathrm{d},\mathcal{V}}^{\alpha\alpha} &\hspace{-1mm} p_{\mathrm{d},\mathcal{V}}^{\alpha\beta} &\hspace{-1mm} p_{\mathrm{d},\mathcal{V}}^{\beta\alpha} &\hspace{-1mm} p_{\mathrm{d},\mathcal{V}}^{\beta\beta} &\hspace{-1mm} p_{\mathrm{d},\mathcal{V}}^{00} &\hspace{-1mm} p_{\mathrm{d},\mathcal{V}}^{0\beta} &\hspace{-1mm} p_{\mathrm{d},\mathcal{V}}^{0\alpha} &\hspace{-1mm} p_{\mathrm{d},\mathcal{V}}^{\beta0} &\hspace{-1mm} p_{\mathrm{d},\mathcal{V}}^{\alpha0}
	\end{bmatrix}^\top.
\end{equation}
It is satisfactory to ascertain that the overall power balance at the nodes corresponds to the power balance at the edges of the topology. In fact, since $i_{\mathrm{dec},\mathcal{G}}^{\lambda_0}$ is always null:
\begin{equation} \label{eq:power_bal_V}
	\sum_{m=1}^{2} \mathbf{p}_{e_{m},\mathcal{V}} = \sum_{n=1}^{3} \mathbf{p}_{v_{n},\mathcal{V}} = 2 p_{\mathrm{d},\mathcal{V}}^{\alpha\alpha} + 6p_{\mathrm{d},\mathcal{V}}^{\beta\beta}
\end{equation}
However, as introduced in \Cref{eq:decoupled_power_basis}, it is possible to determine orthonormal basis to depict the decoupled power patterns. For instance, referring to the edges power variables in \Cref{eq:edge_V_power}, since $\mathrm{rank}(\textbf{p}_{e,\mathcal{V}}) = 2$, a basis $\mathcal{B}_{\mathrm{p}_{e},\mathcal{V}}$ establishes the decoupled power pattern $\textbf{p}_{e,\mathcal{V}}^\mathrm{new}$ as follows:
\begin{equation} \label{eq:V_power_new}
	\mathbf{p}_{e,\mathcal{V}}^\mathrm{new} = \underbrace{\dfrac{1}{\sqrt{2}}\left(\begin{array}{cc}
			1 & -1\\
			1 & 1
		\end{array}\right)}_{\mathcal{B}_{\mathrm{p}_{e},\mathcal{V}}} \mathbf{p}_{e,\mathcal{V}} = \sqrt{2} \left(\begin{array}{cccc}
		0 & -3 & -1 & 0\\
		1 & 0 & 0 & 3
	\end{array}\right) \mathbf{p}_{e,\mathcal{V}}^\mathrm{dec}
\end{equation}
which components inform about the power transfer between the arms and the power transfer between arms and external systems.

\subsection{$\mathcal{D}$ Topology}
The complete three-partite $K_{1,1,1}$ graph topology, commonly named as Delta, is depicted in \Cref{fig:Delta_graph}. The corresponding cell-based converter concept is illustrated in \cite{akagi2011classification}. The incidence matrix $\mathbf{B}_{\mathcal{D}}$ and Laplacian matrix $\mathbf{L}_{\mathcal{D}}$ are:
\begin{equation}
	\mathbf{B}_{\mathcal{D}} = \left(\begin{matrix*}[r]
		-1 & 1 & 0\\
		0 & -1 & 1\\
		1 & 0 & -1
	\end{matrix*}\right) \quad\mathrm{;}\quad 	\mathbf{L}_{\mathcal{D}}= \left(\begin{matrix*}[r]
	2 & -1 & -1\\
	-1 & 2 & -1\\
	-1 & -1 & 2
	\end{matrix*}\right).	
\end{equation}
At the vertices, there exists the following relationship:
\begin{equation}
	\mathbf{L}_{\mathcal{D}}\mathbf{u}_{v,\mathcal{D}} = \mathbf{i}_{v,\mathcal{D}} \quad\text{where}\quad \mathbf{x}_{v,\mathcal{D}} = \begin{bmatrix}
		x_{v_1} & x_{v_2} & x_{v_3}
	\end{bmatrix}^\top.
\end{equation}
Applying \Cref{eq:pure_resistive_network_1} the nodal decoupled variables are achieved:
\begin{equation} \label{eq:decoupled_D}
	\mathbf{M}_{\mathcal{D}} \mathbf{u}_{\mathrm{dec},\mathcal{D}} = \mathbf{i}_{\mathrm{dec},\mathcal{D}} \quad\mathrm{;}\quad \mathbf{M}_{\mathcal{D}} = G_\mathrm{a}\mathrm{diag}\left( 0, 3,3 \right),
\end{equation}
And:
\begin{equation}
	\mathbf{P}_{\mathcal{D}}^{-1} = \dfrac{1}{3}\left(\begin{matrix*}[r]
		1 & 1 & 1\\
		-1 & 2 & -1\\
		-1 & -1 & 2
	\end{matrix*}\right) \quad\mathrm{;}\quad  \mathbf{x}_{\mathrm{dec},\mathcal{D}} = \left(\begin{matrix*}[r]
		x_{\mathrm{dec},\mathcal{D}}^{\lambda_0}\\
		x_{\mathrm{dec},\mathcal{D}}^{\lambda_\alpha}\\
		x_{\mathrm{dec},\mathcal{D}}^{\lambda_\beta}
	\end{matrix*}\right),
\end{equation}
perform the following transformation:
\begin{equation}
	\mathbf{u}_{\mathrm{dec},\mathcal{D}} = \mathbf{P}_{\mathcal{D}}^{-1} \mathbf{u}_{v,\mathcal{D}} \quad\mathrm{;}\quad  \textbf{i}_{\mathrm{dec},\mathcal{D}} = \mathbf{P}_{\mathcal{D}}^{-1} \mathbf{i}_{v,\mathcal{D}}.
\end{equation}
Assuming momentarily that the available internal loop is not employed, from \Cref{eq:edges_variables_graph}, the arm converters variables $\mathbf{x}_{e,\mathcal{D}} = \begin{bmatrix}
	x_{e_1} & x_{e_2} & x_{e_3}
\end{bmatrix}^\top$ are depicted below:
\begin{equation}\label{eq:voltage_modes_D}
	\mathbf{u}_{e,\mathcal{D}} =  \left(\begin{matrix*}[r]
		0 & 2 & 1\\
		0 & -1 & 1\\
		0 & -1 & -2
	\end{matrix*}\right) \mathbf{u}_{\mathrm{dec},\mathcal{D}} 
\end{equation}
and
\begin{equation}\label{eq:current_modes_D}
 \textbf{i}_{e,\mathcal{D}} =  \dfrac{1}{3}\left(\begin{matrix*}[r]
		0 & 2 & 1\\
		0 & -1 & 1\\
		0 & 1 & -2\\
	\end{matrix*}\right) \textbf{i}_{\mathrm{dec},\mathcal{D}}.
\end{equation}
Recalling \Cref{eq:edge_power}, the arm converter instantaneous powers are summarized below:
\begin{equation} \label{eq:D_power_edge}
	\mathbf{p}_{e,\mathcal{D}} = \mathbf{\Gamma}_\mathcal{D}'' \mathbf{p}_{e,\mathcal{D}}^\mathrm{dec} = \dfrac{1}{3}\left(\begin{matrix*}[r]
		4 & 2 & 2 & 1\\
		1 & -1 & -1 & 1\\
		1 & 2 & 2 & 4
	\end{matrix*}\right) \left(\begin{matrix*}[r]
		p_{\mathrm{d},\mathcal{D}}^{\alpha\alpha}\\
		p_{\mathrm{d},\mathcal{D}}^{\alpha\beta}\\
		p_{\mathrm{d},\mathcal{D}}^{\beta\alpha}\\
		p_{\mathrm{d},\mathcal{D}}^{\beta\beta}\\
	\end{matrix*}\right)
\end{equation}
Based on \Cref{eq:node_power}, the nodal powers are defined below:
\begin{equation} \label{eq:D_power_node}
	\mathbf{p}_{v,\mathcal{D}} = 
	\begin{pmatrix*}[r]
		1 & 1 & 1 & 1 & 1 & -1 & -1 & -1 & -1\\
		1 & 0 & 0 & 0 & 1 & 0 & 1 & 0 & 1\\
		0 & 0 & 0 & 1 & 1 & 1 & 0 & 1 & 0
	\end{pmatrix*}
	\mathbf{p}_{v,\mathcal{D}}^\mathrm{dec}
\end{equation}
where:
\begin{equation}
	\mathbf{p}_{v,\mathcal{D}}^\mathrm{dec} = \begin{bmatrix}
		p_{\mathrm{d},\mathcal{D}}^{\alpha\alpha} &\hspace{-1.5mm} p_{\mathrm{d},\mathcal{D}}^{\alpha\beta} &\hspace{-1.5mm} p_{\mathrm{d},\mathcal{D}}^{\beta\alpha} &\hspace{-1.5mm} p_{\mathrm{d},\mathcal{D}}^{\beta\beta} &\hspace{-1.5mm} p_{\mathrm{d},\mathcal{D}}^{00} &\hspace{-1.5mm} p_{\mathrm{d},\mathcal{D}}^{0\beta} &\hspace{-1.5mm} p_{\mathrm{d},\mathcal{D}}^{0\alpha} &\hspace{-1.5mm} p_{\mathrm{d},\mathcal{D}}^{\beta0} &\hspace{-1.5mm} p_{\mathrm{d},\mathcal{D}}^{\alpha0}
	\end{bmatrix}^\top
\end{equation}
then, the overall arms-terminals power balance is satisfied:
\begin{equation} \label{eq:power_bal_D}
	\sum_{m=1}^{3} \mathbf{p}_{e_{m},\mathcal{D}} = \sum_{n=1}^{3} \mathbf{p}_{v_{n},\mathcal{D}} = 2 p_{\mathrm{d},\mathcal{D}}^{\alpha\alpha} + p_{\mathrm{d},\mathcal{D}}^{\alpha\beta} + p_{\mathrm{d},\mathcal{D}}^{\beta\alpha} + 2 p_{\mathrm{d},\mathcal{D}}^{\beta\beta}
\end{equation}
Referring to \Cref{eq:decoupled_power_basis}, since $\mathrm{rank}(\mathbf{p}_{e,\mathcal{D}}) = 3$, a basis $\mathcal{B}_{\mathrm{p}_{e},\mathcal{D}}$ establishes the decoupled power pattern $\mathbf{p}_{e,\mathcal{D}}^\mathrm{new}$ for the Delta topology:
\begin{equation} \label{eq:D_power_new}
	\mathbf{p}_{e,\mathcal{D}}^\mathrm{new} = \mathcal{B}_{\mathrm{p}_{e},\mathcal{D}} \mathbf{p}_{e,\mathcal{D}}  = \dfrac{1}{3}\left(\begin{matrix*}[r]
			2\sqrt{2} & 0 & -1\\
			1/\sqrt{2} & -3/\sqrt{2} & 2\\
			1/\sqrt{2} & 3/\sqrt{2} & 2
		\end{matrix*}\right) \mathbf{p}_{e,\mathcal{D}}
\end{equation}
Therefore, as illustrated in \cite{anton2013elementary}, once an orthonormal basis $\mathcal{B}_{\mathrm{p}}'$ is given, it is always possible to change to another orthonormal basis $\mathcal{B}_{\mathrm{p}}''$ through a proper transition matrix $\mathcal{T}_{\mathcal{B}_{\mathrm{p}}'\rightarrow\mathcal{B}_{\mathrm{p}}''}$ as follows:
\begin{equation} \label{eq:change_basis}
	\mathcal{B}_{\mathrm{p}}'' = \mathcal{T}_{\mathcal{B}_{\mathrm{p}}'\rightarrow\mathcal{B}_{\mathrm{p}}''}  \mathcal{B}_{\mathrm{p}}'.
\end{equation}
This feature can be exploited to get a more physical intuitive power patterns $\mathbf{p}_{e,\mathcal{D}}^\mathrm{new''}$, for instance:
\begin{equation} \label{eq:D_power_new}
	\mathbf{p}_{e,\mathcal{D}}^\mathrm{new''} = \underbrace{\left(\begin{matrix*}[r]
			1 & 1 &1\\
			1 & -1 & 0\\
			1 & 0 & -1
		\end{matrix*}\right)}_{\mathcal{B}_{\mathrm{p}_{e},\mathcal{D}}''} \mathbf{p}_{e,\mathcal{D}} = \left(\begin{matrix*}[r]
		2 & 1 & 1 & 2\\
		1 & 1 & 1 & 0\\
		1 & 0 & 0 & 1
	\end{matrix*}\right)\mathbf{p}_{e,\mathcal{D}}^\mathrm{dec} 
\end{equation}
Moreover, the $\mathcal{D}$ circuit clearly presents a further internal circulating current DOF called $\lambda_\Phi$;  according to \Cref{eq:edges_variables_complete}, the overall arm currents in \Cref{eq:current_modes_D} are rewritten as follows:
\begin{equation}
	\mathbf{i}_{e,\mathcal{D}} =  \dfrac{1}{3}\left(\begin{matrix*}[r]
		0 & 2 & 1 & 3\\
		0 & -1 & 1 & 3\\
		0 & 1 & -2 & 3\\
	\end{matrix*}\right) \underbrace{\left(\begin{matrix*}
			i_\mathrm{dec,\mathcal{D}}^{\lambda_0}\\
			i_\mathrm{dec,\mathcal{D}}^{\lambda_\alpha}\\
			i_\mathrm{dec,\mathcal{D}}^{\lambda_\beta}\\
			i_\mathrm{dec,\mathcal{D}}^{\lambda_\Phi}
		\end{matrix*}\right)}_{\mathbf{i}_{\mathrm{dec},\mathcal{D}}}.
\end{equation}
Assuming that $u_\mathrm{dec,\mathcal{D}}^{\lambda_\Phi} \ll \mathrm{u}_{\mathrm{dec},\mathcal{D}}^{\alpha,\beta}$, the circulating current enables the additional power DOFs depicted below:
\begin{equation} \label{eq:D_power_edge_complete}
	\mathbf{p}_{e,\mathcal{D}} = \dfrac{1}{3}\left(\begin{matrix*}[r]
		4 & 2 & 2 & 1 & 6 & 3\\
		1 & -1 & -1 & 1 & -3 & 3\\
		1 & 2 & 2 & 4 & -3 & -6
	\end{matrix*}\right) \mathbf{p}_{e,\mathcal{D}}^\mathrm{dec} 
\end{equation}
where:
\begin{equation}
	\mathbf{p}_{e,\mathcal{D}}^\mathrm{dec} = \begin{bmatrix}
		p_{\mathrm{d},\mathcal{D}}^{\alpha\alpha} &\hspace{-1.2mm} p_{\mathrm{d},\mathcal{D}}^{\alpha\beta} &\hspace{-1.2mm} p_{\mathrm{d},\mathcal{D}}^{\beta\alpha} &\hspace{-1.2mm} p_{\mathrm{d},\mathcal{D}}^{\beta\beta} &\hspace{-1.2mm} p_{\mathrm{d},\mathcal{D}}^{\Phi \alpha} &\hspace{-1.2mm} p_{\mathrm{d},\mathcal{D}}^{\Phi \beta}
	\end{bmatrix}^\top
\end{equation}
which can be introduced in \Cref{eq:D_power_new} to get the power patterns.\\
\textbf{Remark}: \textit{It is worth to mention that the hypothesis of negligible circulating voltage application $u_\mathrm{dec,\mathcal{D}}^{\lambda_\Phi}$ leads to neglect some mixed product current-voltage that might affect the overall power terms. For instance, the overall power balance depicted in \Cref{eq:power_bal_D}, including $u_\mathrm{dec,\mathcal{D}}^{\lambda_\Phi}$, can be rewritten as follows:
\begin{equation} 
	\sum_{m=1}^{3} \mathbf{p}_{e_{m},\mathcal{D}} = 2 p_{\mathrm{dec},\mathcal{D}}^{\alpha\alpha} + p_{\mathrm{dec},\mathcal{D}}^{\alpha\beta} + p_{\mathrm{dec},\mathcal{D}}^{\beta\alpha} + 2 p_{\mathrm{dec},\mathcal{D}}^{\beta\beta} + 3 p_{\mathrm{dec},\mathcal{D}}^{\Phi\Phi}.
\end{equation}
Where, the term $p_{\mathrm{dec},\mathcal{D}}^{\Phi\Phi}$, depending on the arm passive parameters $L_\mathrm{a}$ and $G_\mathrm{a}$, determines both the reactive power absorbed by the internal reactance and the power losses occurring by employing the circulating current DOF.}\\
From now on, for the topologies presenting the internal circulating current loops, the corresponding voltages is considered negligible compared with the nodal voltage DOFs values.

\subsection{2$\mathcal{V}$ Topology}
The complete bi-partite $K_{2,2}$ graph topology is depicted in \Cref{fig:2V_graph}. Mostly known as Wheatstone bridge circuit \cite{wheatstone1843account}, proposed in \cite{dey2017modular} as a cell-based converter concept. The incidence matrix $\mathbf{B}_{2\mathcal{V}}$ and Laplacian matrix $\mathbf{L}_{2\mathcal{V}}$ are presented below:
\begin{equation}
	\mathbf{B}_{2\mathcal{V}} = \left(\begin{matrix*}[r]
		-1 & 1 & 0 & 0\\
		0 & -1 & 1 & 0\\
		0 & 0 & -1 & 1\\
		1 & 0 & 0 & -1
	\end{matrix*}\right)
\end{equation}
and
\begin{equation}
	 \mathbf{L}_{2\mathcal{V}}= \left(\begin{matrix*}[r]
		2 & -1 & 0 & -1\\
		-1 & 2 & -1 & 0\\
		0 & -1 & 2 & -1\\
		-1 & 0 & -1 & 2
	\end{matrix*}\right).
\end{equation}
At the vertices, there exists the following relationship:
\begin{equation}
	\mathbf{L}_{2\mathcal{V}}\mathbf{u}_{v,2\mathcal{V}} = \mathbf{i}_{v,2\mathcal{V}} \quad\text{where}\quad \mathbf{x}_{v,2\mathcal{V}} = \begin{bmatrix}
		x_{v_1} & x_{v_2} & x_{v_3} & x_{v_4}
	\end{bmatrix}^\top.
\end{equation}
Recalling \Cref{eq:pure_resistive_network_1}, the nodal decoupled variables are achieved:
\begin{equation} \label{eq:decoupled_2V}
	\mathbf{M}_{2\mathcal{V}} \mathbf{u}_{\mathrm{dec},2\mathcal{V}} = \mathbf{i}_{\mathrm{dec},2\mathcal{V}} \quad\mathrm{;}\quad \mathbf{M}_{2\mathcal{V}} = G_\mathrm{a}\mathrm{diag}\left(0,4,2,2 \right).
\end{equation}
And
\begin{equation}
	\mathbf{P}_{2\mathcal{V}}^{-1} = \dfrac{1}{4}\left(\begin{matrix*}[r]
		1 & 1 & 1 & 1\\
		-1 & 1 & -1 & 1\\
		-2 & 0 & 2 & 0\\
		0 & -2 & 0 & 2
	\end{matrix*}\right) \quad\text{;}\quad \mathbf{x}_{\mathrm{dec},2\mathcal{V}} = \left(\begin{matrix*}[r]
	x_{\mathrm{dec},\mathrm{2\mathcal{V}}}^{\lambda_0}\\
	x_{\mathrm{dec},\mathrm{2\mathcal{V}}}^{\lambda_\alpha}\\
	x_{\mathrm{dec},\mathrm{2\mathcal{V}}}^{\lambda_\beta}\\
	x_{\mathrm{dec},\mathrm{2\mathcal{V}}}^{\lambda_\gamma}
	\end{matrix*}\right).
\end{equation}
perform the following transformation:
\begin{equation}
	\mathbf{u}_{\mathrm{dec},\mathrm{2\mathcal{V}}} = \mathbf{P}_{\mathrm{2\mathcal{V}}}^{-1} \mathbf{u}_{v,\mathrm{2\mathcal{V}}} \quad\mathrm{;}\quad  \mathbf{i}_{\mathrm{dec},\mathrm{2\mathcal{V}}} = \mathbf{P}_{\mathrm{2\mathcal{V}}}^{-1} \mathbf{i}_{v,\mathrm{2\mathcal{V}}}.
\end{equation}
\textbf{Remark}: \textit{It is then confirmed that the $2\mathcal{V}$ based converter allows two decoupled modes, $\lambda_\beta$ and $\lambda_\gamma$, which involve two separated subsets of terminals ($V_1 = \left\lbrace v_1,v_3\right\rbrace $ and $V_2 = \left\lbrace v_2,v_4\right\rbrace $). This fundamental property enables two decoupled ports, and it was experimentally proved through the Wheatstone bridge}.\\
Assuming, momentarily, that the unique internal loop is not employed, from \Cref{eq:edges_variables_graph}, the arm converters $\mathbf{x}_{e,2\mathcal{V}} = \begin{bmatrix}
	x_{e_1} & x_{e_2} & x_{e_3} & x_{e_4}
\end{bmatrix}^\top$ variables are depicted below:
\begin{equation}\label{eq:voltage_modes_2V}
	\mathbf{u}_{e,2\mathcal{V}} =  \left(\begin{matrix*}[r]
		0 & 2 & 1 & -1\\
		0 & -2 & 1 & 1\\
		0 & 2 & -1 & 1\\
		0 & -2 & -1 & -1
	\end{matrix*}\right) \mathbf{u}_{\mathrm{dec},2\mathcal{V}} 
\end{equation}
and
\begin{equation}\label{eq:current_modes_2V}
	\mathbf{i}_{e,2\mathcal{V}} =  \dfrac{1}{2}\left(\begin{matrix*}[r]
		0 & 1 & 1 & -1\\
		0 & -1 & 1 & 1\\
		0 & 1 & -1 & 1\\
		0 & -1 & -1 & -1
	\end{matrix*}\right) \mathbf{i}_{\mathrm{dec},2\mathcal{V}}.
\end{equation}
Again, from \Cref{eq:edge_power}, arm converter powers are depicted:
\begin{equation} \label{eq:H_power_edge}
	\mathbf{p}_{e,2\mathcal{V}} = \dfrac{1}{2}\left(\begin{matrix*}[r]
		2 & 1 & 2 & -1 & 1 & -2 & -1 & -1 & 1\\
		2 & -1 & -2 & -1 &1 & -2 & 1 & 1& 1\\
		2 & -1 & -2 & 1& 1 & 2 & -1& -1& 1\\
		2 & 1 & 2 & 1 & 1 & 2 & 1 & 1 & 1
	\end{matrix*}\right) \mathbf{p}_{e,2\mathcal{V}}^\mathrm{dec}
\end{equation}
where:
\begin{equation}
	\small\mathbf{p}_{e,2\mathcal{V}}^\mathrm{dec} = \begin{bmatrix}
		p_{\mathrm{d},2\mathcal{V}}^{\alpha\alpha} &\hspace{-1.5mm} p_{\mathrm{d},2\mathcal{V}}^{\alpha\beta} &\hspace{-1.5mm} p_{\mathrm{d},2\mathcal{V}}^{\beta\alpha} &\hspace{-1.5mm} p_{\mathrm{d},2\mathcal{V}}^{\alpha\gamma} &\hspace{-1.5mm} p_{\mathrm{d},2\mathcal{V}}^{\beta\beta} &\hspace{-1.5mm} p_{\mathrm{d},2\mathcal{V}}^{\gamma\alpha} &\hspace{-1.5mm} p_{\mathrm{d},2\mathcal{V}}^{\beta\gamma} &\hspace{-1.5mm} p_{\mathrm{d},2\mathcal{V}}^{\gamma\beta} &\hspace{-1.5mm} p_{\mathrm{d},2\mathcal{V}}^{\gamma\gamma}
	\end{bmatrix}^\top.
\end{equation}
Based on \Cref{eq:node_power}, it is possible to achieve the following nodal instantaneous power relationship:
\begin{equation} \label{eq:2V_powernode}
	\mathbf{p}_{v,2\mathcal{V}} = \mathbf{\Gamma}_{4\times 16}' \mathbf{p}_{v,2\mathcal{V}}.
\end{equation}
Once again, it is proved that the instantaneous power balance among the terminals and arms coincides:
 \begin{equation} \label{eq:power_bal_2V}
 	\sum_{m=1}^{4} \mathbf{p}_{e_{m},2\mathcal{V}} = \sum_{n=1}^{4} \mathbf{p}_{v_{n},2\mathcal{V}} = 4 p_{\mathrm{dec},2\mathcal{V}}^{\alpha\alpha} + 2p_{\mathrm{dec},2\mathcal{V}}^{\beta\beta} + 2p_{\mathrm{dec},2\mathcal{V}}^{\gamma\gamma}.
 \end{equation}
Referring to \Cref{eq:decoupled_power_basis}, since $\mathrm{rank}(\mathbf{p}_{e,2\mathcal{V}}) = 4$, a basis $\mathcal{B}_{\mathrm{p}_{e},2\mathcal{V}}$ establishes the following decoupled power pattern $\mathbf{p}_{e,2\mathcal{V}}^\mathrm{new}$:
\begin{equation} \label{eq:2V_power_new}
	\mathbf{p}_{e,2\mathcal{V}}^\mathrm{new} = \mathcal{B}_{\mathrm{p}_{e},2\mathcal{V}} \mathbf{p}_{e,2\mathcal{V}} = \dfrac{1}{2}\left(\begin{matrix*}[r]
		1 & 1 & 1 & 1\\
		1 & -1 & -1 & 1\\
		-1 & -1 & 1 & 1\\
		-1 & 1 & -1 & 1\\
		\end{matrix*}\right) \mathbf{p}_{e,2\mathcal{V}}.
\end{equation}
As presented for the $\mathcal{D}$ topology, in the $2\mathcal{V}$ graph, an internal circulating current mode $\lambda_\Phi$ is available. Accordingly to this, the arms converter currents in \Cref{eq:current_modes_2V} are expressed below:
\begin{equation}
	\mathbf{i}_{e,2\mathcal{V}} =  \dfrac{1}{2}\left(\begin{matrix*}[r]
		0 & 1 & 1 & -1 & 2\\
		0 & -1 & 1 & 1 & 2\\
		0 & 1 & -1 & 1 & 2\\
		0 & -1 & -1 & -1 & 2
	\end{matrix*}\right) \left(\begin{matrix*}
			i_\mathrm{dec,2\mathcal{V}}^{\lambda_0}\\
			i_\mathrm{dec,2\mathcal{V}}^{\lambda_\alpha}\\
			i_\mathrm{dec,2\mathcal{V}}^{\lambda_\beta}\\
			i_\mathrm{dec,2\mathcal{V}}^{\lambda_\gamma}\\
			i_\mathrm{dec,2\mathcal{V}}^{\lambda_\Phi}
		\end{matrix*}\right),
\end{equation}
which reflects on decoupled edges power patterns as follows:
\begin{equation} \label{eq:2V_power_edge_complete}
	\begin{cases}
		&	p_{e,2\mathcal{V}}^\mathrm{new,1} = 2p_{\mathrm{dec},2\mathcal{V}}^{\alpha\alpha} + p_{\mathrm{dec},2\mathcal{V}}^{\beta\beta} + p_{\mathrm{dec},2\mathcal{V}}^{\gamma\gamma}\\
		&	p_{e,2\mathcal{V}}^\mathrm{new,2} = p_{\mathrm{dec},2\mathcal{V}}^{\alpha\beta} - 2p_{\mathrm{dec},2\mathcal{V}}^{\Phi\gamma} + 2p_{\mathrm{dec},2\mathcal{V}}^{\beta\alpha}\\
		&	p_{e,2\mathcal{V}}^\mathrm{new,3} = p_{\mathrm{dec},2\mathcal{V}}^{\alpha\gamma} - 2p_{\mathrm{dec},2\mathcal{V}}^{\Phi\beta} + 2p_{\mathrm{dec},2\mathcal{V}}^{\gamma\alpha}\\
		&	p_{e,2\mathcal{V}}^\mathrm{new,4} = p_{\mathrm{dec},2\mathcal{V}}^{\beta\gamma} - 4p_{\mathrm{dec},2\mathcal{V}}^{\Phi\alpha} + p_{\mathrm{dec},2\mathcal{V}}^{\gamma\beta}\\
	\end{cases}
\end{equation}
\textbf{Remark}: \textit{Though the circulating current DOF does not play any role on the overall converter power balance, its main purpose is to enable additional power terms which are responsible of the power transfer among the internal arms}.
\vspace{-4mm}
\subsection{$\mathcal{Y}$ Topology}
The common $\mathcal{Y}$ graph topology in \Cref{fig:Y_graph} is a complete bi-partite graphs $K_{3,1}$. A Cell-based converter structure can be encountered in \cite{akagi2011classification}. $\mathbf{B}_\mathcal{Y}$ and $\mathbf{L}_\mathcal{Y}$ are depicted below:
\begin{equation}
	\mathbf{B}_\mathcal{Y} = \left(\begin{matrix*}[r]
		-1 & 1 & 0 & 0\\
		-1 & 0 & 1 & 0\\
		-1 & 0 & 0 & 1
	\end{matrix*}\right) \quad;\quad \mathbf{L}_\mathcal{Y}= \left(\begin{matrix*}[r]
	3 & -1 & -1 & -1\\
	-1 & 1 & 0 & 0\\
	-1 & 0 & 1 & 0\\
	-1 & 0 & 0 & 1
	\end{matrix*}\right)
\end{equation}
At the vertices, there exists the following relationship:
\begin{equation}
	\mathbf{L}_{\mathcal{Y}}\mathbf{u}_{v,\mathcal{Y}} = \mathbf{i}_{v,\mathcal{Y}} \quad\text{where}\quad \mathbf{x}_{v,\mathcal{Y}} = \begin{bmatrix}
		x_{v_1} & x_{v_2} & x_{v_3} & x_{v_4}
	\end{bmatrix}^\top.
\end{equation}
Recalling \Cref{eq:pure_resistive_network_1}, the nodal decoupled variables are achieved:
\begin{equation} \label{eq:decoupled_Y}
	\mathbf{M}_{\mathcal{Y}} \mathbf{u}_{\mathrm{dec},\mathcal{Y}} = \mathbf{i}_{\mathrm{dec},\mathcal{Y}} \quad\mathrm{;}\quad \mathbf{M}_{\mathcal{Y}} = G_\mathrm{a}\mathrm{diag}\left(0,4,1,1 \right).
\end{equation}
Decoupled modes are expressed by:
\begin{equation}\label{eq:modes_Y}
	\mathbf{u}_{\mathrm{dec},\mathrm{\mathcal{Y}}} = \mathbf{P}_{\mathrm{\mathcal{Y}}}^{-1} \mathbf{u}_{v,\mathrm{\mathcal{Y}}} \quad\mathrm{;}\quad  \mathbf{i}_{\mathrm{dec},\mathrm{\mathcal{Y}}} = \mathbf{P}_{\mathrm{\mathcal{Y}}}^{-1} \mathbf{i}_{v,\mathrm{\mathcal{Y}}}.
\end{equation}
where:
\begin{equation}
	\mathbf{P}_\mathcal{Y}^{-1} = \dfrac{1}{4}\left(\begin{matrix*}[r]
		1 & 1 & 1 & 1\\
		-1 & 1/3 & 1/3 & 1/3\\
		0 & -4/3 & 8/3 & -4/3\\
		0 & -4/3 & -4/3 & 8/3
	\end{matrix*}\right)
\end{equation}
and
\begin{equation}
 \mathbf{x}_{\mathrm{dec},\mathcal{Y}} = \begin{bmatrix}
 	x_{\mathrm{dec},\mathrm{2\mathcal{Y}}}^{\lambda_0} & x_{\mathrm{dec},\mathrm{2\mathcal{Y}}}^{\lambda_\alpha} & x_{\mathrm{dec},\mathrm{2\mathcal{Y}}}^{\lambda_\beta} & x_{\mathrm{dec},\mathrm{2\mathcal{Y}}}^{\lambda_\gamma}
 \end{bmatrix}^\top.
\end{equation}
From \Cref{eq:edges_variables_graph}, the arm converters variable $\mathbf{x}_{e,\mathcal{Y}} = \begin{bmatrix}
	x_{e_1} & x_{e_2} & x_{e_3}
\end{bmatrix}^\top$ are depicted below:
\begin{equation}
	\mathbf{u}_{e,\mathcal{Y}} =  \left(\begin{matrix*}[r]
		0 & 4 & -1 & -1\\
		0 & 4 & 1 & 0\\
		0 & 4 & 0 & 1
	\end{matrix*}\right) \mathbf{i}_{\mathrm{dec},\mathcal{Y}},
\end{equation}
\begin{equation}
	\mathbf{i}_{e,\mathcal{Y}} = \left(\begin{matrix*}[r]
		0 & 1 & -1 & -1\\
		0 & 1 & 1 & 0\\
		0 & 1 & 0 & 1
	\end{matrix*}\right) \mathbf{i}_{\mathrm{dec},\mathcal{Y}}.
\end{equation}
So, the arm converter instantaneous powers are:
\begin{equation} \label{eq:Y_power_edge}
	\mathbf{p}_{e,\mathcal{Y}} = \small\left(\begin{matrix*}[r]
		4 & -1 & -4 & -1 & 1 & -4 & 1 & 1 & 1\\
		4 & 1 & 4 & 0 & 1 & 0 & 0 & 0& 0\\
		4 & 0 & 0 & 1 & 0 & 4 & 0 & 0 & 1
	\end{matrix*}\right) \mathbf{p}_{e,\mathcal{Y}}^\mathrm{dec}
\end{equation}
where:
\begin{equation}
	\small\mathbf{p}_{e,\mathcal{Y}}^\mathrm{dec} = \begin{bmatrix}
		p_{\mathrm{d},\mathcal{Y}}^{\alpha\alpha} &\hspace{-1mm} p_{\mathrm{d},\mathcal{Y}}^{\alpha\beta} &\hspace{-1mm} p_{\mathrm{d},\mathcal{Y}}^{\beta\alpha} &\hspace{-1mm} p_{\mathrm{d},\mathcal{Y}}^{\alpha\gamma} &\hspace{-1mm} p_{\mathrm{d},\mathcal{Y}}^{\beta\beta} &\hspace{-1mm} p_{\mathrm{d},\mathcal{Y}}^{\gamma\alpha} &\hspace{-1mm} p_{\mathrm{d},\mathcal{Y}}^{\beta\gamma} &\hspace{-1mm} p_{\mathrm{d},\mathcal{Y}}^{\gamma\beta} &\hspace{-1mm} p_{\mathrm{d},\mathcal{Y}}^{\gamma\gamma}
	\end{bmatrix}^\top.
\end{equation}
Referring to \Cref{eq:decoupled_power_basis}, since $\mathrm{rank}(\mathbf{p}_{e,\mathcal{Y}}) = 3$, the orthonormal basis $\mathcal{B}_{\mathrm{p}_{e},\mathcal{Y}}$ establishes the following decoupled power pattern $\mathbf{p}_{e,\mathcal{Y}}^\mathrm{new}$:
\begin{equation} \label{eq:Y_power_new}
	\mathbf{p}_{e,\mathcal{Y}}^\mathrm{new} =\mathcal{B}_{\mathrm{p}_{e},\mathcal{Y}} \mathbf{p}_{e,\mathcal{Y}} = \dfrac{1}{3}\left(\begin{matrix*}[r]
			1 & 1 & 1\\
			1 & -1 & -1\\
			-1 & -1 & 1\\
		\end{matrix*}\right) \mathbf{p}_{e,\mathcal{Y}},
\end{equation}
which can be written substituting $\mathbf{p}_{e,\mathcal{Y}}$ from the \Cref{eq:Y_power_edge}.

\subsubsection{Transition matrix to Clarke transformation}
Assuming to fix node $v_\mathrm{1} = 0$, and eliminating mode $\lambda_0$ appearing in $\mathbf{P}_\mathcal{Y}^{-1}$; let $\mathbf{C}$ be the Clarke transformation depicted in \Cref{fig:general_transf_1}, it is possible to change the basis through a proper transition matrix $\mathcal{T}_{\mathbf{C} \rightarrow\mathbf{P}_\mathcal{Y}^{-1}}$ as follows:
\begin{equation} \label{eq:change_basis_Clarke}
	\mathbf{C} = \mathcal{T}_{\mathbf{C} \rightarrow\mathbf{P}_\mathcal{Y}^{-1}}  \mathbf{P}_\mathcal{Y}^{-1} = \mathcal{T}_{\mathbf{C} \rightarrow\mathbf{P}_\mathcal{Y}^{-1}} \dfrac{1}{4}\left(\begin{matrix*}[r]
			1/3 & 1/3 & 1/3\\
			-4/3 & 8/3 & -4/3\\
			-4/3 & -4/3 & 8/3\\
		\end{matrix*}\right);
\end{equation}
then, it results:
\begin{equation} \label{eq:change_basis_Clarke_2}
	\small\mathbf{C} = \underbrace{\dfrac{1}{3}\left(\begin{matrix*}[r]
			0 & -1 & -1\\
			\dfrac{4\left(\sqrt{3} - \sqrt{2} \right) }{3} & \dfrac{\sqrt{3}}{3} & \dfrac{-\sqrt{2}}{3}\\
			4 & 0 & 0\\
		\end{matrix*}\right)}_{\mathcal{T}_{\mathbf{C} \rightarrow\mathbf{P}_\mathcal{Y}^{-1}}}  \mathbf{P}_\mathcal{Y}^{-1};
\end{equation}
\vspace{-10mm}
\subsection{2$\mathcal{Y}$ Topology}\label{ssec:2Y_analysis}
\begin{figure}[!h]
	\centering
	\includegraphics[]{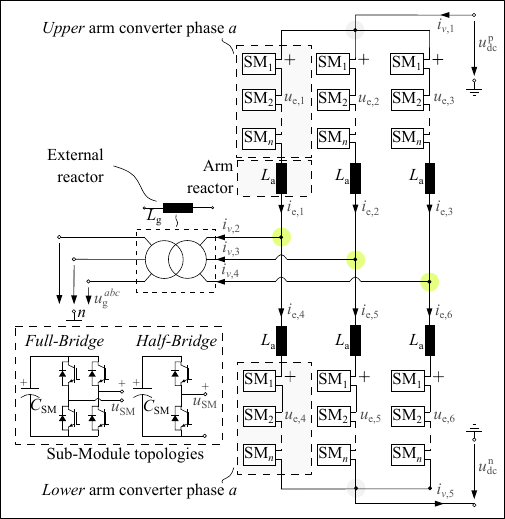}
	\label{fig:edge_partition}
	\caption{AC/DC Modular Multilevel Converter schematic.}
	\label{fig:MMC_converter}
\end{figure}
The complete bi-partite $K_{3,2}$ graph topology is depicted in \Cref{fig:2Y_graph}. The topology is mostly known as MMC \cite{lesnicar2003innovative} and the AC/DC MMC electrical drawing is depicted in \Cref{fig:MMC_converter}. $\mathbf{B}_{2\mathcal{Y}}$ and $\mathbf{L}_{2\mathcal{Y}}$ are presented below:
\begin{equation}
	\mathbf{B}_{2\mathcal{Y}} = \left(\begin{matrix*}[r]
		-1 & 1 & 0 & 0 & 0\\
		-1 & 0 & 1 & 0 & 0\\
		-1 & 0 & 0 & 1 & 0\\
		0 & -1 & 0 & 0 & 1\\
		0 & 0 & -1 & 0 & 1\\
		0 & 0 & 0 & -1 & 1
	\end{matrix*}\right);
\end{equation}
and
\begin{equation}
	\mathbf{L}_{2\mathcal{Y}} = \left(\begin{matrix*}[r]
		3 & -1 & -1 & -1 & 0\\
		-1 & 2 & 0 & 0 & -1\\
		-1 & 0 & 2 & 0 & -1\\
		-1 & 0 & 0 & 2 & -1\\
		0 & -1 & -1 & -1 & 3
	\end{matrix*}\right).
\end{equation}
At the vertices, there exists the following relationship:
\begin{equation}
	\mathbf{L}_{2\mathcal{Y}}\mathbf{u}_{v,2\mathcal{Y}} = \mathbf{i}_{v,2\mathcal{Y}} \quad\text{;}\quad \mathbf{x}_{v,2\mathcal{Y}} = \begin{bmatrix}
		x_{v_1} & x_{v_2} & x_{v_3} & x_{v_4} & x_{v_5}
	\end{bmatrix}^\top.
\end{equation}
Recalling \Cref{eq:pure_resistive_network_1}, the nodal decoupled variables are achieved:
\begin{equation} \label{eq:decoupled_2Y}
	\mathbf{M}_{2\mathcal{Y}} \mathbf{u}_{\mathrm{dec},2\mathcal{Y}} = \mathbf{i}_{\mathrm{dec},2\mathcal{Y}} \quad\mathrm{;}\quad \mathbf{M}_{2\mathcal{Y}} = G_\mathrm{a}\mathrm{diag}\left(0,3,5,2,2 \right).
\end{equation}
The nodal normal modes are expressed by:
\begin{equation}\label{eq:modes_2Y}
	\mathbf{u}_{\mathrm{dec},2{\mathcal{Y}}} = \mathbf{P}_{2{\mathcal{Y}}}^{-1} \mathbf{u}_{v,2{\mathcal{Y}}} \quad\mathrm{;}\quad  \mathbf{i}_{\mathrm{dec},2{\mathcal{Y}}} = \mathbf{P}_{2{\mathcal{Y}}}^{-1} \mathbf{i}_{v,2{\mathcal{Y}}}.
\end{equation}
where:
\begin{equation}
	\mathbf{P}_{2\mathcal{Y}}^{-1} = \dfrac{1}{5}\left(\begin{matrix*}[r]
		1 & 1 & 1 & 1 & 1\\
		-5/2 & 0 & 0 & 0 & 5/2\\
		3/2 & -1 & -1 & -1 & 3/2\\
		0 & -5/3 & 10/3 & -5/3 & 0\\
		0 & -5/3 & -5/3 & 10/3 & 0
	\end{matrix*}\right)
\end{equation}
and
\begin{equation}
	\mathbf{x}_{\mathrm{dec},2\mathcal{Y}} = \begin{bmatrix}
		x_{\mathrm{dec},\mathrm{2\mathcal{Y}}}^{\lambda_0} & x_{\mathrm{dec},\mathrm{2\mathcal{Y}}}^{\lambda_\alpha} & x_{\mathrm{dec},\mathrm{2\mathcal{Y}}}^{\lambda_\beta} & x_{\mathrm{dec},\mathrm{2\mathcal{Y}}}^{\lambda_\gamma} & x_{\mathrm{dec},\mathrm{2\mathcal{Y}}}^{\lambda_\delta}
	\end{bmatrix}^\top
\end{equation}
\textbf{Remark}: \textit{It is confirmed that the mode $\lambda_\alpha$ and the modes $\lambda_\gamma-\lambda_\delta$ involve two separated subsets of terminals: $V_1 = \left\lbrace v_1,v_5\right\rbrace $ and $V_2 = \left\lbrace v_2,v_3,v_4\right\rbrace $. This fundamental property enables two decoupled ports composed by two-terminals and three-terminals respectively}.\\
Assuming momentarily that the internal current loops are not employed, from \Cref{eq:edges_variables_graph}, the arm converters variables $\mathbf{x}_{e,2\mathcal{Y}} = \begin{bmatrix}
	x_{e_1} & x_{e_2} & x_{e_3} & x_{e_4} & x_{e_5} & x_{e_6}
\end{bmatrix}^\top$  variables are depicted below:
\begin{equation}
	\mathbf{u}_{e,2\mathcal{Y}} =  \left(\begin{matrix*}[r]
		0 & 1 & -5/3 & -1 & -1\\
		0 & 1 & -5/3 & 1 & 0\\
		0 & 1 & -5/3 & 0 & 1\\
		0 & 1 & 5/3 & 1 & 1\\
		0 & 1 & 5/3 & -1 & 0\\
		0 & 1 & 5/3 & 0 & -1
	\end{matrix*}\right) \mathbf{u}_{\mathrm{dec},2\mathcal{Y}},
\end{equation}
\begin{equation}
	\mathbf{i}_{e,2\mathcal{Y}} =  \dfrac{1}{3}\left(\begin{matrix*}[r]
		0 & 1 & -1 & -3/2 & -3/2\\
		0 & 1 & -1 & 3/2 & 0\\
		0 & 1 & -1 & 0 & 3/2\\
		0 & 1 & 1 & 3/2 & 3/2\\
		0 & 1 & 1 & -3/2 & 0\\
		0 & 1 & 1 & 0 & -3/2
	\end{matrix*}\right) \mathbf{i}_{\mathrm{dec},2\mathcal{Y}}.
\end{equation}
As for the previous topologies, referring to \Cref{eq:decoupled_power_basis}, the orthonormal basis $\mathcal{B}_{\mathrm{p}_{e},2\mathcal{Y}}$ establishes the following decoupled power pattern $\mathbf{p}_{e,2\mathcal{Y}}^\mathrm{new}$:
\begin{equation} \label{eq:Y_power_new}
	\mathbf{p}_{e,2\mathcal{Y}}^\mathrm{new} = \underbrace{\dfrac{1}{6}\left(\begin{matrix*}[r]
			\sqrt{6} & \sqrt{6} & \sqrt{6} & \sqrt{6} & \sqrt{6} & \sqrt{6}\\
			-\sqrt{6} & -\sqrt{6} & -\sqrt{6} & \sqrt{6} & \sqrt{6} & \sqrt{6}\\
			-3 & 3 & 0 & 3 & -3 & 0\\
			3 & -3 & 0 & 3 & -3 & 0\\
			\sqrt{3} & \sqrt{3} & -2\sqrt{3} & \sqrt{3} & \sqrt{3} & -2\sqrt{3}\\
			-\sqrt{3} & -\sqrt{3} & 2\sqrt{3} & \sqrt{3} & \sqrt{3} & -2\sqrt{3}\\
		\end{matrix*}\right)}_{\mathcal{B}_{\mathrm{p}_{e},2\mathcal{Y}}} \mathbf{p}_{e,2\mathcal{Y}}
\end{equation}
\subsubsection{Circulating current loops} 
Looking at the topology, from \Cref{eq:int_dec_description}, two independent current loops are encountered:
\begin{equation} \label{eq:2Y_loops}
	\mathbf{i}_{\mathrm{dec},2\mathcal{Y}}^\Phi = \left(\begin{matrix*}[r]
		i_{\mathrm{dec},2\mathcal{Y}}^{\Phi_1}\\
		i_{\mathrm{dec},2\mathcal{Y}}^{\Phi_2}
	\end{matrix*}\right) = \left(\begin{matrix*}[r]
	1 & 0 & -1 & 1 & 0 & -1\\
	1 & -1 & 0 & 1 & -1 & 0\\
	\end{matrix*}\right) \mathbf{i}_{e,2\mathcal{Y}},
\end{equation}
the overall decoupled power patterns are summarized below:
\begin{equation}\label{eq:2Y_Pnew1}
\small	\begin{split}
	p_{e,2\mathcal{Y}}^\mathrm{new,1} &= \dfrac{\sqrt{6}}{18} \left( 6p_{\mathrm{dec},2\mathcal{Y}}^{\alpha\alpha} + 10p_{\mathrm{dec},2\mathcal{Y}}^{\beta\beta} + 6p_{\mathrm{dec},2\mathcal{Y}}^{\gamma\gamma} + 3p_{\mathrm{dec},2\mathcal{Y}}^{\gamma\delta} +\right.\\ & 
	\left. + 3p_{\mathrm{dec},2\mathcal{Y}}^{\delta\gamma} + 6p_{\mathrm{dec},2\mathcal{Y}}^{\delta\delta}\right); 
	\end{split}
\end{equation}
\vspace{-2mm}
\begin{equation}\label{eq:2Y_Pnew2}
	\small\begin{split}
		p_{e,2\mathcal{Y}}^\mathrm{new,2} &= \dfrac{\sqrt{6}}{9} \left( 5p_{\mathrm{dec},2\mathcal{Y}}^{\alpha\beta} + 3p_{\mathrm{dec},2\mathcal{Y}}^{\beta\alpha} + 3p_{\mathrm{dec},2\mathcal{Y}}^{\Psi_{1}\gamma} + 6p_{\mathrm{dec},2\mathcal{Y}}^{\Psi_{1}\delta} +\right.\\ & 
		\left. + 6p_{\mathrm{dec},2\mathcal{Y}}^{\Psi_{2}\gamma} + 3p_{\mathrm{dec},2\mathcal{Y}}^{\Psi_{2}\delta}\right); 
	\end{split}
\end{equation}
\vspace{-2mm}
\begin{equation}\label{eq:2Y_Pnew3}
	\small\begin{split}
		p_{e,2\mathcal{Y}}^\mathrm{new,3} &= \dfrac{2p_{\mathrm{dec},2\mathcal{Y}}^{\alpha\gamma}}{3} + p_{\mathrm{dec},2\mathcal{Y}}^{\gamma\alpha} + \dfrac{p_{\mathrm{dec},2\mathcal{Y}}^{\alpha\delta}}{3} + \dfrac{p_{\mathrm{dec},2\mathcal{Y}}^{\delta\alpha}}{2} + \dfrac{5p_{\mathrm{dec},2\mathcal{Y}}^{\Psi_{1}\beta}}{3} \\ & 
		+ p_{\mathrm{dec},2\mathcal{Y}}^{\Psi_{1}\gamma} + \dfrac{10p_{\mathrm{dec},2\mathcal{Y}}^{\Psi_{2}\beta}}{3} + p_{\mathrm{dec},2\mathcal{Y}}^{\Psi_{1}\delta} + p_{\mathrm{dec},2\mathcal{Y}}^{\Psi_{2}\delta}; 
	\end{split}
\end{equation}
\vspace{-2mm}
\begin{equation}\label{eq:2Y_Pnew4}
	\small\begin{split}
		p_{e,2\mathcal{Y}}^\mathrm{new,4} &= \dfrac{2p_{\mathrm{dec},2\mathcal{Y}}^{\beta\gamma}}{3} + \dfrac{5p_{\mathrm{dec},2\mathcal{Y}}^{\gamma\beta}}{3} + \dfrac{p_{\mathrm{dec},2\mathcal{Y}}^{\beta\delta}}{3} + \dfrac{5p_{\mathrm{dec},2\mathcal{Y}}^{\delta\beta}}{6} + \dfrac{p_{\mathrm{dec},2\mathcal{Y}}^{\gamma\delta}}{2} \\ & 
		+ \dfrac{p_{\mathrm{dec},2\mathcal{Y}}^{\delta\gamma}}{2} + \dfrac{p_{\mathrm{dec},2\mathcal{Y}}^{\delta\delta}}{2} + p_{\mathrm{dec},2\mathcal{Y}}^{\Psi_{1}\alpha} + 2p_{\mathrm{dec},2\mathcal{Y}}^{\Psi_{2}\alpha}; 
	\end{split}
\end{equation}
\vspace{-2mm}
\begin{equation}\label{eq:2Y_Pnew5}
\small	\begin{split}
		p_{e,2\mathcal{Y}}^\mathrm{new,5} &= \dfrac{\sqrt{3}}{6} \left( 2p_{\mathrm{dec},2\mathcal{Y}}^{\beta\delta} + 2p_{\mathrm{dec},2\mathcal{Y}}^{\gamma\gamma} + 5p_{\mathrm{dec},2\mathcal{Y}}^{\delta\beta} + p_{\mathrm{dec},2\mathcal{Y}}^{\gamma\delta} +\right.\\ & 
		\left. + p_{\mathrm{dec},2\mathcal{Y}}^{\delta\gamma} - p_{\mathrm{dec},2\mathcal{Y}}^{\delta\delta} + 6p_{\mathrm{dec},2\mathcal{Y}}^{\Psi_{1}\alpha}\right); 
	\end{split}
\end{equation}
\vspace{-2mm}
\begin{equation}\label{eq:2Y_Pnew6}
	\small	\begin{split}
		p_{e,2\mathcal{Y}}^\mathrm{new,6} &= \dfrac{\sqrt{3}}{6} \left( 2p_{\mathrm{dec},2\mathcal{Y}}^{\alpha\delta} + 3p_{\mathrm{dec},2\mathcal{Y}}^{\delta\alpha} + 10p_{\mathrm{dec},2\mathcal{Y}}^{\Psi_{1}\beta} + 2p_{\mathrm{dec},2\mathcal{Y}}^{\Psi_{1}\gamma} -\right.\\ & 
		\left. - 2p_{\mathrm{dec},2\mathcal{Y}}^{\Psi_{1}\delta} + 4p_{\mathrm{dec},2\mathcal{Y}}^{\alpha\gamma} + 2p_{\mathrm{dec},2\mathcal{Y}}^{\alpha\delta}\right). 
	\end{split}
\end{equation}
\textbf{Remark}: \textit{As for the $\mathcal{D}$ and $2\mathcal{V}$ topologies, the circulating current DOFs main purpose is to enable additional power terms which are responsible of the power transfer among the internal arms. On this regard, both the AC and DC domain can be applied simultaneously to $i_{\mathrm{dec},2\mathcal{Y}}^{\Phi_1}$ and $i_{\mathrm{dec},2\mathcal{Y}}^{\Phi_2}$. By playing in amplitude, phase and frequency, it is possible to activate and deactivate specific power terms}.
\subsubsection{Nodal Normal Modes Validation}
\begin{figure}[!h]
	\centering
	\subfloat[]
	{
		\includegraphics[]{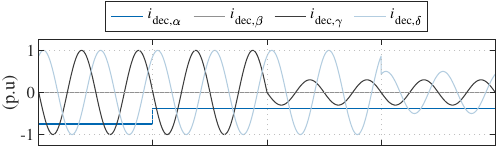}
		\label{fig:norm}
	}\\
	\vspace{-3.5mm}
	\subfloat[]
	{
		\includegraphics[]{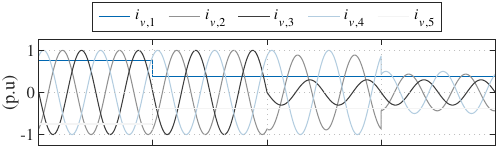}
		\label{fig:nodes}
	}\\
	\vspace{-3.5mm}
	\subfloat[]
	{
		\includegraphics[]{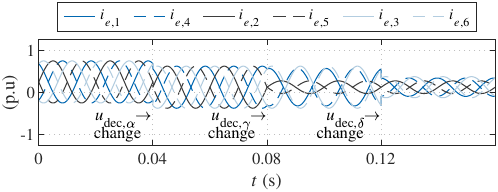}
		\label{fig:edge}
	}
	\vspace{-2.5mm}
	\caption{In (a) the topology current DOFs and their step changes behaviour. In (b) and (c) the current variables for the terminals and for the arms respectively based on the DOFs.}
	\label{fig:validation}
\end{figure}

Referring to a $2\mathcal{Y}$ pure resistive network topology, a simulation is performed to prove the effective decoupling among the above mentioned nodal normal modes. The results depicted in \Cref{fig:validation} demonstrate that the variation of the nodal normal modes current variables, \Cref{fig:norm}, reflects to both the terminals and the edges currents of the topology, (\Cref{fig:nodes} and \Cref{fig:edge}), while any current interaction between the DOFs is verified. The test is performed by assuming the voltage normal nodal mode $u_{\text{dec},\alpha}$ to be purely DC, with a step change at 0.04 s; the voltage nodal normal mode $u_{\text{dec},\beta}$ is set to be zero. The voltage nodal normal modes, involving the three terminals connected to the three-phase system, $u_{\text{dec},\gamma}$ and $u_{\text{dec},\delta}$ are purely AC (50 Hz) and, step changes, at 0.08 as and 0.12 s respectively, are performed.
\vspace{-2mm}
\vspace{-2mm}
\section{Further Insights From the Analysis}\label{sec:insights}
\vspace{0mm}

\subsection{Complete $k$-partite Graphs Inspiring Further $k$-port Converter Topologies Families}
\begin{figure}[!h]
	\centering
	\subfloat[]
	{
	\includegraphics[]{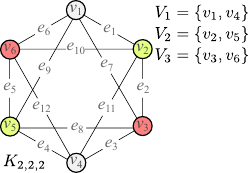}
	\label{fig:222}
	}
	\subfloat[]
	{
	\includegraphics[]{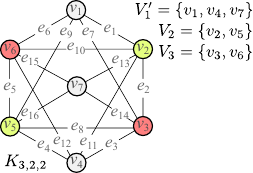}
	\label{fig:322}
	}
	\hspace{2mm}
\vspace{-2mm}
	\caption{Graph representation of possible multi-decoupled-port converter families based on complete three-partite graphs: A $K_{2,2,2}$ graph in (a) and a $K_{3,2,2}$ in (b).}
	\label{fig:further_multi_port}
\end{figure}

The methodology illustrated along the manuscript can be additionally employed for inspiring alternative cell-based converter arrangements accommodating, for instance, more than two ports. In fact, as the complete bi-partite topology $K_{3,2}$, analysed in \Cref{ssec:2Y_analysis}, provides two decoupled ports formed by three and two terminals respectively; the complete three-partite graphs $K_{2,2,2}$ and $K_{3,2,2}$, depicted in \Cref{fig:222} and \Cref{fig:322} respectively, might be proposed to build-up three-decoupled-port converters topologies. In particular, from \Cref{fig:222}, the topology could interact with three external single-phase-based systems independently through the three terminals subsets $V_1$,$V_2$ and $V_3$. From \Cref{fig:322}, the topology could interact with two external single-phase-based systems and an external three-phase-based system independently through the three vertices subsets $V_1'$,$V_2$ and $V_3$.
\vspace{-6mm}

Limiting the analysis to the nodal DOFs, the graph Laplacian eigenvector of the $K_{2,2,2}$ topology are depicted below:
\begin{equation}
	\small	\mathbf{P}_{K_{2,2,2}}^{-1} = \dfrac{1}{6}\left(\begin{matrix*}[r]
			1 & 1 & 1 & 1 & 1 & 1\\
			-1 & 2 & -1 & -1 & 2 & 1\\
			-1 & -1 & 2 & -1 & -1 & 2\\
			-3 & 0 & 0 & 3 & 0 & 0\\
			0 & -3 & 0 & 0 & 3 & 0\\
			0 & 0 & -3 & 0 & 0 & 3\\		
		\end{matrix*}\right),
\end{equation}
where:
\begin{equation}
\mathbf{P}_{K_{2,2,2}}^{-1} = \begin{bmatrix}
	\vec{\lambda}_0 & \vec{\lambda}_\alpha & \vec{\lambda}_\beta & \vec{\lambda}_\gamma & \vec{\lambda}_\delta & \vec{\lambda}_\epsilon
\end{bmatrix}^\top.
\end{equation}
Screening $\mathbf{P}_{K_{2,2,2}}^{-1}$, it is then confirmed that the normal modes $\vec{\lambda}_\gamma$, $\vec{\lambda}_\delta$ and $\vec{\lambda}_\epsilon$ involve three different decoupled subsets of terminals, or ports, $(v_1 - v_4)$, $(v_2 - v_5)$ and $(v_3 - v_6)$ respectively. Then, if properly controlled, the arm voltage sources might consequently apply the three decoupled voltage DOFs to regulate current among the three interconnected systems independently. \\
On the other hand, the graph Laplacian eigenvector of the $K_{3,2,2}$ topology are evaluated as follows:
\begin{equation}
	\small\mathbf{P}_{K_{3,2,2}}^{-1} = \dfrac{1}{7}\left(\begin{matrix*}[r]
		1 & 1 & 1 & 1 & 1 & 1 & 0\\
		0 & -7/2 & 0 & 0 & 7/2 & 0 & 0\\
		0 & 0 & -7/2 & 0 & 0 & 7/2 & 0\\
		-1 & -1 & 5/2 & -1 & -1 & 5/2 & 1\\
		4/3 & -1 & -1 & 4/3 & -1 & -1 & 4/3\\
		-7/3 & 0 & 0 & 14/3 & 0 & 0 & -7/3\\
		-7/3 & 0 & 0 & -7/3 & 0 & 0 & 14/3\\
	\end{matrix*}\right)
\end{equation}
where:
\begin{equation}
	\mathbf{P}_{K_{3,2,2}}^{-1} = \begin{bmatrix}
		\vec{\lambda}_0 & \vec{\lambda}_\alpha & \vec{\lambda}_\beta & \vec{\lambda}_\gamma & \vec{\lambda}_\delta & \vec{\lambda}_\epsilon & \vec{\lambda}_\zeta
	\end{bmatrix}^\top.
\end{equation}
Screening $\mathbf{P}_{K_{3,2,2}}^{-1}$, it is then confirmed that the normal modes $\vec{\lambda}_\alpha$ and $\vec{\lambda}_\beta$ involve two different decoupled subsets of terminals, $V_2$ and $V_3$ respectively. The normal modes $\vec{\lambda}_\epsilon$ and $\vec{\lambda}_\zeta$ involve another subset of terminals $V_1'$. Then, if properly controlled, the arm voltage sources might consequently apply the three decoupled voltage DOFs to regulate current among the three interconnected systems independently. 

\subsection{Ports Galvanic Isolation \textit{vs} Ports Current Decoupling}
\begin{figure}[!h]
\centering
\subfloat[]
{
	\includegraphics[]{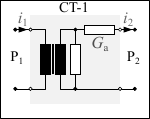}
	\label{fig:t}
}
\subfloat[]
{
	\includegraphics[]{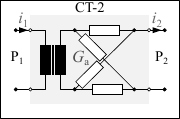}
	\label{fig:r}
}
\subfloat[]
{
	\includegraphics[]{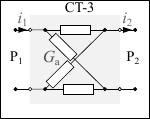}
	\label{fig:h}
}\\
\vspace{-3mm}
\subfloat[]
{
	\includegraphics[]{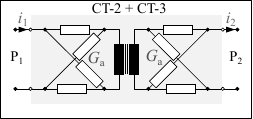}
	\label{fig:w}
}
\subfloat[]
{
	\includegraphics[]{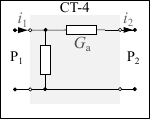}
	\label{fig:q}
}
\caption{Resistive-based case scenarios referred to \Cref*{tab:galvanic_is_decoupled_currents}.}
\label{fig:galv_is_dec_curr}
\end{figure}

The ports current decoupling is not to be confused with the ports galvanic isolation attribute. In general, a network providing galvanic isolated ports does not mean that the corresponding port currents are inherently decoupled to each other, and vice-versa. In fact, as a two port transformer principle of operation is based on the magnetic coupling between primary and secondary side, the two corresponding currents are not actually disjointed to each other \cite{kulkarni2004transformer}. With respect to this, topologies can be further categorized by both properties: \textit{Galvanic Isolated Ports} (GIP) based topologies and \textit{Decoupled Durrents Ports} (DCP) based topologies. \Cref{tab:galvanic_is_decoupled_currents}, together with the representative conductance based circuits in \Cref{fig:galv_is_dec_curr}, summarizes the possible topologies scenarios (CT).\\
Referring to a pure resistive sub-network, as experienced in \cite{wheatstone1843account}, the network topology might act as a current firewall between the connected ports. This inherent decoupling property can be further investigated for applications demanding for strong decoupling among the interconnected systems either in normal and contingency operation. In cell-based converter topologies, in practice, this features can be further exploited for the the inherent port fault blocking capability of the topology: being the ports currents decoupled to each other, a fault current through a short circuit at one port, can be extinguished by only disabling the corresponding decoupled voltage DOF involving that specific port. However, it is worth to mention that the fault blocking capability of the converter will be practically enabled by the SM technology installed within the arm, \Cref{fig:real_edge}, as illustrated in \cite{yao2019modular}.\\
Furthermore, for the complete $k$-partite based topologies, another aspect to be highlighted is the capability to interconnect, by the sub-network converter, multiple systems with different voltage amplitudes, frequencies and pole configurations without further intermediate energy conversion stages. For instance, taking the $K_{2,2}$ based topology in \Cref{fig:h}, it can be possible to connect an asymmetrical monopole DC at the port $P_1$, to a symmetrical monopole at the $P_2$; this cannot be practicable for the topology case in \Cref{fig:q}, for instance.
\begin{table}[h!]\footnotesize
	\renewcommand{\arraystretch}{1}
	\caption{Galvanic isolation \textit{vs} current decoupling.}\label{tab:galvanic_is_decoupled_currents}
	\centering
	\begin{tabularx}{0.3\textwidth}{ c  c  c  c  c }
		\hline
		\hline
		& CT-1  & CT-2  & CT-3  & CT-4  \\	
		\hline
		GIP 		& \cmark & \cmark& \xmark & \xmark \\ 
		\hline
		DCP		& \xmark & \cmark & \cmark & \xmark \\
		\hline
		\hline
	\end{tabularx}
	\end{table}

\subsection{Analogies With Mechanical Mass-Spring Systems}	
\begin{figure}[!h]
	\centering
	\subfloat[]
	{
	\includegraphics[]{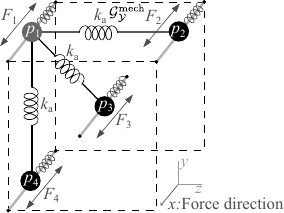}
	\label{fig:m}
	}
	\hspace*{-0em}
	\subfloat[]
	{
	\includegraphics[]{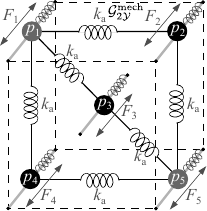}
	\label{fig:n}
	}
	\caption{Mass-spring systems representation of the $\mathcal{Y}$ (a) and $2\mathcal{Y}$ (b) converter networks.}
	\label{fig:electrical_mech_analogy}
\end{figure}

The fundamental comprehension on the converter degrees of freedom can be aided by the mechanical mass-spring system analogy. In \Cref{fig:electrical_mech_analogy}, the complete bi-partite $K_{1,3}$ and $K_{3,2}$ mechanical based graphs are illustrated. Edges resistances are replaced by springs with the same stiffness constant $K_\mathrm{a}$ and nodes are replaced by ideal masses which are interconnected to each other through the spring network. The masses position and forces direction applied to mass is assumed to be along the axis $x$. Recalling \Cref{eq:nodal_relation} for the electrical network, for the mechanical system the Hooke's law in \Cref{eq:mech_el} has to be analysed.
\begin{equation} \label{eq:mech_el}
		\mathbf{L}_{\mathcal{G}}^{\mathrm{mech}} \mathbf{F}_{v,\mathcal{G}}^\mathrm{mech} = -\mathbf{x}_{v,\mathcal{G}}^{\mathrm{mech}}\\
\end{equation}
where $\mathbf{L}_{\mathcal{G}}^{\mathrm{mech}} = \mathbf{B}_\mathcal{G}^\top \mathbf{W}_\mathcal{G}^{\mathrm{mech}}  \mathbf{B}_\mathcal{G}$, $\mathbf{F}_{v,\mathcal{G}}^\mathrm{mech}$ is the vector of nodal forces and $\mathbf{x}_{v,\mathcal{G}}^{\mathrm{mech}}$ indicates the nodal positions vector. Matrix $\mathbf{W}_\mathcal{G}^{\mathrm{mech}} = G_\mathrm{a} \mathbf{I}_{m}$ where $G_\mathrm{a} = 1/k_\mathrm{a}$ represents the stiffness matrix of the system. Analogously to the electrical network, the mechanical system nodal degrees of freedom can be achieved by looking at the graph Laplacian eigenvector-eigenvalues. In the specific cases, recalling \Cref{eq:nodes_modes}, \Cref{eq:decoupled_Y} and \Cref{eq:decoupled_2Y}, decoupled mechanical force patterns are expressed by:
\begin{equation}
	\mathbf{F}_{\mathrm{dec},\mathrm{\mathcal{Y}}}^{\mathrm{mech}} = \mathbf{P}_{\mathrm{\mathcal{Y}}}^{-1} \mathbf{F}_{v,\mathrm{\mathcal{Y}}}^{\mathrm{mech}} \quad\mathrm{;}\quad  \mathbf{F}_{\mathrm{dec},\mathrm{2\mathcal{Y}}}^{\mathrm{mech}} = \mathbf{P}_{\mathrm{2\mathcal{Y}}}^{-1} \mathbf{F}_{v,\mathrm{\mathcal{Y}}}^{\mathrm{mech}}.
\end{equation}

\vspace{-2mm}
\section{Conclusion}\label{sec:conclusion}
Aiming to get a more comprehensive and systematic ordering among the cell-based converter topologies degrees of freedom, the manuscript proposed a further analysis methodology perspective. The presented approach is rather linked to both circuit and graph analysis. In particular, the graph Laplacian eigenvector-eigenvalue based assessment is founded to be clearly revealing the decoupled voltage-current modes of a certain graph-based converter at the points of connection with external systems. Throughout the manuscript, several existing converter topologies have been addressed by a step-by-step approach to characterize their fundamental degrees of freedom. Both the nodal and edges (arm converters) voltage-current and power variables are then described from the decoupled reference frame variables. 
Furthermore, eventual internal circulating current loops of the topology are included to fully depict all the decoupled edges power terms expressions. The manuscript concludes providing some further points of discussion related to the presented methodology; for instance,  the intriguing potential to determine decoupled subset of terminals, which can be exploited to unfold multi-decoupled-port converter topologies.

%

\appendices



\ifCLASSOPTIONcaptionsoff
  \newpage
\fi
\vspace{-2mm}
\bibliography{refs_2}
\bibliographystyle{IEEEtran}\small
\vspace{-1cm}

\end{document}